\documentclass[onecolumn,sort&compress,numbers]{els-mrw} 

\usepackage{amsmath,amssymb,bm,amsfonts,amsthm,makeidx,graphicx}
\usepackage{txfonts}
\usepackage{helvet}
\usepackage{epstopdf}
\usepackage{xfrac}
\usepackage{phoenician}
\usepackage{caption}
\usepackage{footmisc}

\begin{document}

\section{The QCD Running Coupling}

~

\author[1]{Alexandre~Deur}%

\address[1]{\orgname{Thomas Jefferson National Accelerator Facility}, \orgdiv{Physics Division}, \orgaddress{Newport News, VA 23606, USA}}


\maketitle

\begin{abstract}[Abstract]
We describe the coupling of the strong force. 
Denoted as $\alpha_s$, it sets the strength of that force, just as $G$ or $\alpha$ specify the strength of gravity and electromagnetism. 
Its value depends on the scale at which phenomena are observed.
In this chapter, we will explain the nature of the coupling, the quantum origin of its scale dependence, and the crucial consequences this entails for quantum chromodynamics,
the gauge theory for the strong force.
We describe the theories for the calculation of $\alpha_s$, using the perturbative method at high-momentum scales (equivalently, short-distance scales) 
and nonperturbative approaches at low-momentum scales (equivalently, long-distance scales).
We also present the experimental determinations of $\alpha_s$ at both short and long distance scales.
\end{abstract}

\begin{keywords}
 	Strong Force\sep QCD\sep running coupling \sep perturbative \sep nonperturbative
\end{keywords}

\begin{figure}[h]
	\centering
	\includegraphics[width=0.5\textwidth]{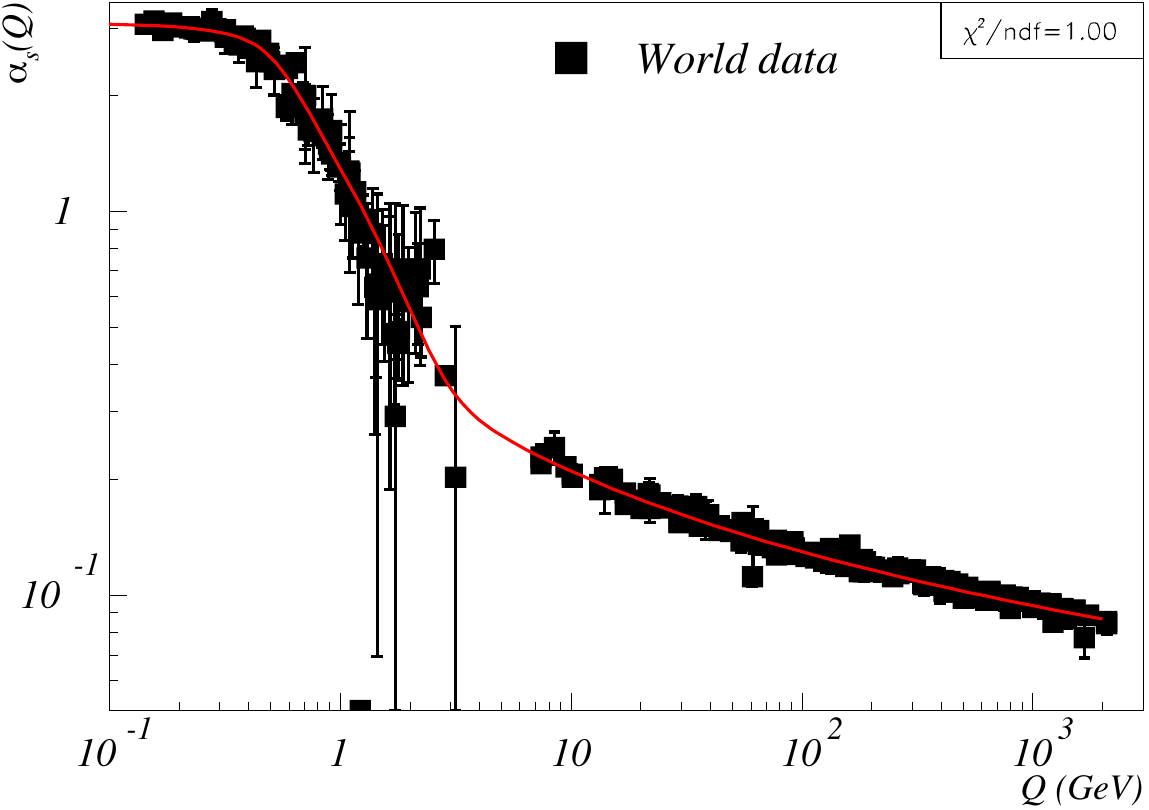}
	\vspace*{-41ex}  
        \begin{center} \hspace*{3cm} 
        {\LARGE \color{red}{---}}~~$aT_r/\ln\big(\frac{Q^2+Q^2_r}{\Lambda^2}\big)$\\ \hspace*{3.cm} 	
        \end{center}
        \vspace*{30ex} 
	\caption{The QCD coupling, $\alpha_s$, in function of momentum scale $Q$. 
	The points are experimental data. The red line is simple fit
	using the 1-loop perturbative formula $\propto 1/\ln\big(\sfrac{Q^2}{\Lambda^2}\big)$,
	a Fermi-Dirac function for the scale $Q_r$=$b/\big(e^{(Q^2-c)/d} +1\big)$ regularizing the divergence at $Q=\Lambda$, 
	and another Fermi-Dirac function $T_r$=$\big(1+(\pi-1)/(e^{(Q-f)/g}+1)\big)$ to smooth the low-to-high $Q$ transition. The parameters values yielding
	$\chi^2/n.d.f.=1.00$ are: $a=1.56$, $\Lambda=0.246$~GeV, $b=0.808$~GeV, $c=0.11$~GeV$^2$, $d=0.20$~GeV$^2$, $f=$1.29~GeV and $g=$0.59~GeV.}
	\label{fig:titlepage}
\end{figure}

\begin{glossary}[Nomenclature]
	\begin{tabular}{@{}lp{34pc}@{}}
AdS/CFT & anti-de Sitter/conformal field theory \\
AdS/QCD & anti-de Sitter/QCD \\
CEBAF & Continuous Electron Beam Accelerator Facility \\
CERN & European Organization for Nuclear Research\\
CLAS & CEBAF Large Acceptance spectrometer \\
CSR & commensurate scale relations \\
DESY & Deutsches Elektronen-Synchrotron \\
DGLAP & Dokshitzer--Gribov--Lipatov--Altarelli--Parisi \\
DIS & deep inelastic scattering \\
DSE & Dyson-Schwinger equation \\
EIC & electron ion collider (at Brookhaven National Laboratory) \\
EicC & electron ion collider in China \\
FLAG & flavor lattice averaging group \\
HERA & Hadron-Electron Ring Accelerator \\
HLFQCD & holographic light-front QCD \\
IR & infrared (physics/phenomena) \\
JLab & Thomas Jefferson National Accelerator Facility \\
LEP & Large Electron-Positron collider \\
LF & Light Front \\
LGT & lattice gauge theory \\
LHC & large hadron collider (at CERN) \\
LO & leading order \\
MOM (scheme) & momentum space subtraction (scheme) \\
MS (and $\overline{\rm MS}$) & Minimal Subtraction \\
NLO (and N$^i$LO) & next to ... leading order \\
PDG & Particle Data Group and associated publications \\
pQCD & perturbative QCD \\
QCD & quantum chromodynamics \\
QED & quantum electrodynamics \\
QFT & quantum field theory \\
RGE & renormalization group equation \\
RS & Renormalization Scheme \\
STI & Slavnov-Taylor Identities \\
SU(N) & special unitary (group of degree n) \\
UV & ultraviolet (physics/phenomena) \\

	\end{tabular}
\end{glossary}

\section*{Objectives}
\begin{itemize}
	\item Origin of the running of the couplings specifying the strength of forces.
	\item Behavior of the QCD coupling $\alpha_s$ at high-energy, {\it viz} in the QCD perturbative domain. 
	\item Behavior of $\alpha_s$ at low-energy, {\it viz} in the QCD nonperturbative domain. 
	\item Methods to determines $\alpha_s$ from experiments or theories.  
\end{itemize}

\subsection{Introduction and Overview}\label{overview}
The strong force
, which binds quarks together into hadrons, and nucleons into nuclei, is one of the four known 
basic interactions of nature. The three others are electromagnetism
, the weak force
, and gravity
. A crucial attribute of forces is their magnitude, that is, how strongly they act at a given distance and for a given amount of matter. 
This magnitude is set by the {\it coupling constant}, {\it e.g.}, for gravity, Newton's constant $G$. The strong force coupling constant is named $\alpha_s$.
The quantum field theory (QFT) of the strong force is quantum chromodynamics (QCD). 
Just like the source of the electromagnetism is the electric charge, the QCD sources are the {\it color charges}. 
They come in three types: red, green and blue, in contrast to electromagnetism with its single type of electric charge. 
Color charges are carried by both quarks and gluons, again unlike in electromagnetism where photons are electrically neutral.
Quarks carry one color, anti-quarks one anti-color (anti-red, anti-green or anti-blue), 
and gluons, one color and one anti-color. Then, according to QCD, the strong force is the effect felt by quarks when they 
trade colors {\it via} gluons, with $\alpha_s$ giving the gluon emission probability. Since gluons are also colored, they 
interact among themselves, with $\alpha_s$ playing the same r\^ole as in the quark case.

In classical physics, a force coupling {\it constant} is indeed a fixed number. This is not so in the quantum realm. From symmetry viewpoint, 
the cause is the appearance of a {\it quantum anomaly}. 
From the process viewpoint, the cause is quantum loops, these ephemeral materialization and subsequent annihilation of pairs of particle-antiparticle. 
This happens as follows: classically, a fundamental force between two bodies obeys the inverse-square law. However, once matter is scrutinized at short distances where
quantum effects become important, quantum loop  effects start to be felt. 
This causes the force to deviate from the inverse-square law and makes QFT calculations diverge, requiring a {\it renormalization program}
. It is natural in the renormalization process to preserve the 
inverse-square law and assign to coupling constants the extra distance-dependence from quantum loops, thereby making the hitherto constant couplings 
scale-dependent. Henceforth, we will thus refer to them simply as {\it couplings}, while their acquired scale dependence is referred to as their {\it running}. 

For the electric force, the running of the coupling, $\alpha$, is small: from macroscopic distances to the shortest ones probed in high-energy physics, 
$\alpha$ increases by about 10\% of its value. 
In contrast, the running is large for $\alpha_s$, its value changing by several folds. Another crucial difference with $\alpha$ is that while the latter increases as 
distances decrease, $\alpha_s$ does the opposite: it decreases and even vanishes in the zero-distance limit, see Fig.~\ref{fig:titlepage}. 
This phenomenon, known as quark {\it asymptotic freedom}, is crucial since a small $\alpha_s$ makes the powerful method of perturbation theory available. 
This availability for QCD at short distances means that the phenomenology of $\alpha_s$ is well understood there. In practice,
challenges and uncertainties remain stemming from the facts that 
({\bf A}) high-order perturbative QCD (pQCD) calculations are difficult due to the non-abelian nature of QCD that makes it non-linear; and 
({\bf B}) pQCD provides the running of $\alpha_s$ but not its absolute magnitude. The latter is determined experimentally or using 
nonperturbative techniques. The consequence of ({\bf A}) and ({\bf B}) is that significant theoretical and experimental uncertainties are
associated with $\alpha_s$. For comparison, $\alpha$ is known with a precision of about 1 part in 10 billion, while for $\alpha_s$ at short distance, 
it is currently about 1\%, {\it viz} $10^8$ times worse. This notwithstanding, the nature of $\alpha_s$ and its evolution is well understood in the 
pQCD domain.
Another crucial r\^ole of $\alpha_s$ 
is as the expansion parameter for pQCD calculations. 
Hence, knowing $\alpha_s$ precisely is essential for achieving the accuracy demanded by high-energy scattering experiments, {\it e.g.},
those at CERN's LHC, which test the Standard Model 
and explore its expected extension
. Currently, a sub-percent precision on $\alpha_s$ is required so that this uncertainty does not dominate other ones. 
Such precision is presently barely achieved ($\Delta \alpha_s/\alpha_s = 0.85\%$) by combining the world data on $\alpha_s$~\cite{Workman:2022ynf}.
This explains why measuring $\alpha_s$ at short distances remains an active and important sector of research.

At distances similar to hadron sizes, $\alpha_s$ becomes too large to allow for perturbation theory. 
This has historically clouded our understanding of $\alpha_s$ at long-distance, with predictions on the latter 
ranging from it vanishing again to being infinite. It was eventually realized that to good approximation, $\alpha_s$  is
measurable in that domain. The data were followed by several nonperturbative predictions consistent with the data. 
All this revealed that at long distances, $\alpha_s$ stops running and is relatively large (Fig.~\ref{fig:titlepage}). 
Knowing $\alpha_s$ at long distances allowed for successful predictions of many basic hadronic quantities. 
The ability to conduct such calculations cannot be overstated since, in nature, most phenomena involving QCD are nonperturbative. 
For example, nonperturbative QCD produces almost all of the universe visible mass.

\subsection{Force couplings, and why they run}
\label{running couplings}

Classically, the magnitude of a force is characterized by its {\it coupling constant}, a universal factor that quantifies the force between 
two static elementary sources. The sources are the elementary electric charge $e$ for electromagnetism, 
the color charge $g$ for QCD, and the weak isospin $g_w$ for the weak force
. For gravity, there is presently no known elementary 
mass; therefore, its coupling constant $G$ has a 1/mass$^2$ dimension (in natural units, $\hbar=1=c$). In contrast, $\alpha$, $\alpha_s$, 
and the weak coupling $\alpha_w$ are dimensionless. 
Accordingly, and except gravity, a coupling is by definition proportional to the elementary charge squared:
$\alpha \equiv e^2/4\pi$,
$\alpha_s \equiv g^2/4\pi$ and
$\alpha_w \equiv g_w^2/4\pi$. (With the electroweak unification
, the weak coupling can be expressed in terms of $\alpha$ and the Weinberg angle $\theta_w$: $\alpha_w = \alpha/\sin^2\theta_W$.)

\begin{figure}[t!]
\centering
\includegraphics[width=1\textwidth]{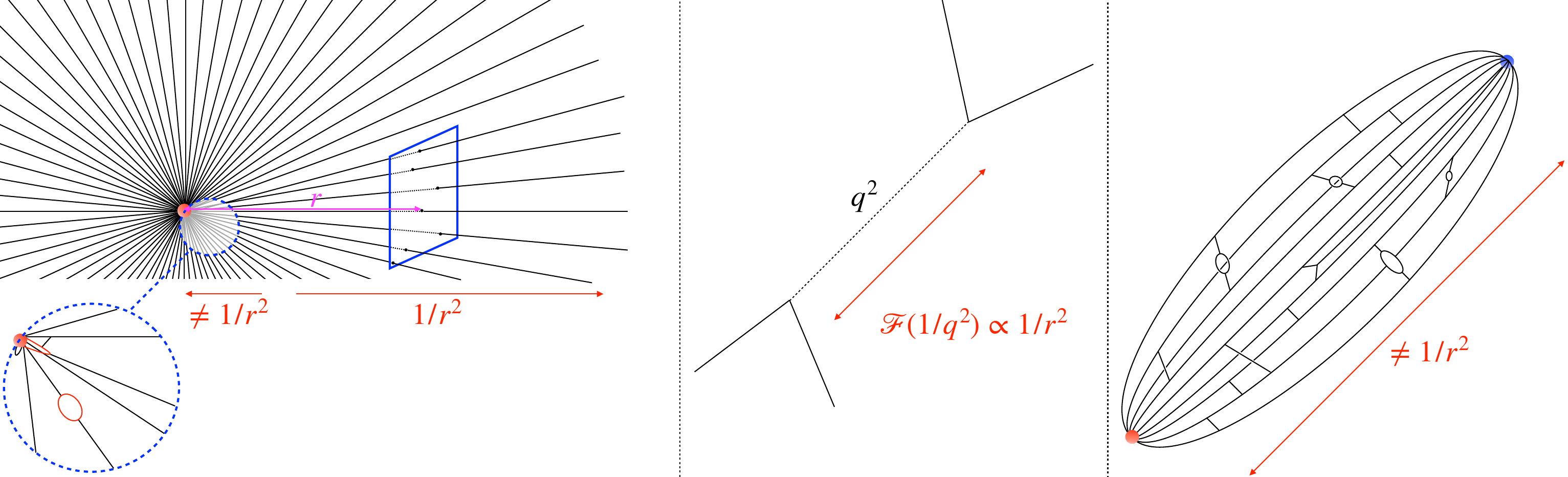}
\caption{\textbf{The inverse square law of forces, and where it fails.} For static sources, a classical force follows
$\vec{F}=\frac{\mathcal{A} c_1 c_2}{r^2}e^{-m|\vec{r}|}(1+m|\vec{r}|) \vec{u}_r$, where 
$\mathcal{A}$ is the force coupling, 
$c_1$ and $c_2$ are the source charges (masses for gravity, electric charge for electricity, color for QCD, and weak isospin for the weak force), 
$\vec r$ is the vector linking the two sources, 
$\vec{u}_r\equiv \vec{r}/|\vec{r}|$, and
$m$ is the field mass, set to $m=0$ in this chapter. 
Classically, the inverse square law, {\it i.e.} the $1/r^2$ factor, stems from the force flux that freely spreads in 3D 
space (left panel). The straight black lines are the field lines, the blue square represents the unit area, 
distant from the source by $r$, through which the flux is measured. 
The force equals the flux going through the area. 
The red sphere symbolizes one of the sources (the other is the test particle and therefore not shown). 
In QFT, the $1/r^2$ dependence originates from the Fourier transform of the gauge boson propagator (dashed line in middle panel) 
in the Born approximation; $1/r^2 \propto\mathcal F(1/q^2)$ with $q$ the boson 4-momentum. 
The $1/r^2$ law fails at short distances because of additional $r-$dependences from quantum loops, 
as illustrated in the magnified area in the left panel. The straight lines now depict the trajectories of the gauge bosons rather than field lines. 
Fermion loops are shown in red. 
The $1/r^2$ law also fails for strongly interacting non-linear theories because the self-interaction of the field prevents its free spreading ({\it viz} propagation) in space. QCD is the prototypical example of such phenomenon (right panel). 
In the renormalization process, the additional scale dependence from quantum loops is folded into the definition of the coupling rather than modifying the gauge boson propagator, thereby making the coupling scale-dependent. This is the origin of its running.
}
\label{inverse-square-law}
\end{figure}
 For linear theories, {\it i.e.}, those for which the field superposition principle holds, and for static sources emitting a massless field, 
 the coupling is the magnitude-setting factor $\mathcal{A}$ that relates the force to the charges $c_1$ and $c_2$ of two sources 
 divided by $r^2$, where $r$ is the source separation: 
 $F =\frac{\mathcal{A} c_1 c_2}{r^2}$, see Fig.~\ref{inverse-square-law}. 
 The $1/r^2$ 
 law was first understood by Faraday as the dilution of the force flux as it uniformly expands through 3D 
 space. In QFT, the $1/r^2$ law stems from the propagator of the massless gauge boson carrying the force in the Born approximation {\it viz}, 
 by one-boson exchange, the leading order (LO) in perturbation theory. In momentum space, this yields the familiar propagator form 
 $\propto 1/q^2$, where $q$ is the boson 4-momentum.\footnote{
 When considering processes that involve spacelike momenta, we will use $Q^2=-q^2 >0 $.}
Faraday's picture agrees with the QFT interpretation since a propagator expresses the particle's probability to travel from here to there, and the gauge bosons are emitted isotropically from their source.
Yet, when scrutinized in detail, the Born approximation fails: higher orders in perturbation are necessary and alter the $1/r^2$ law. 
In particular, higher-order diagrams containing loops of particles seemingly yield infinities. 
Those are managed by regularization and renormalization
, wherein the additional 
scale-dependence caused by the loops is folded into the coupling, making it scale dependent. This makes clear two important facts:
({\bf A}) the running of a coupling is purely a quantum phenomenon, due to higher-order loop diagrams contributing to the force magnitude;
({\bf B}) the choice to fold the additional scale dependence into the coupling is a matter of convenience and preference.\footnote{
In fact, this choice is not always made. One traditionally publishes experimental lepton scattering data corrected 
so that they are expressed in the Born approximation. This entails performing {\it radiative corrections}~\cite{Mo:1968cg} that include the loops 
that make $\alpha$ to run. 
Therefore, the classical coupling {\it constant} value, $\alpha \approx 1/137$, is used in the calculations to extract results from lepton scattering 
experiments rather than the running coupling.}

\begin{figure}[t!]
\centering
\includegraphics[width=0.5\textwidth]{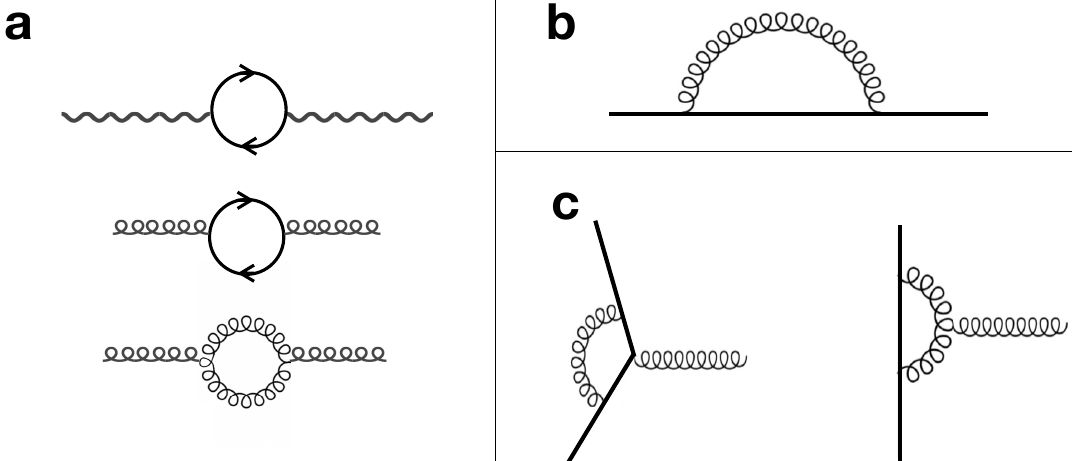}
\caption{\label{quantum_effects} 
\textbf{Short distance QED (panel a-top graph) and QCD (all panels) processes making $\bm \alpha$ and $\bm {\alpha_s}$ to run}.
(a): Vacuum polarization;
(b): Quark self-energy;
(c): Vertex corrections.
Other graphs exist, see~\cite{Muta:1998vi} for the list up to next-to-leading order (NLO).} 
\end{figure}

In quantum electrodynamics (QED), vacuum polarization (Fig.~\ref{quantum_effects}a-upper graph) is the only effect that causes the coupling to run. 
In QCD, more processes contribute: 
vacuum polarization (Fig.~\ref{quantum_effects}a-lower graphs),
quark self-energy (Fig.~\ref{quantum_effects}b), 
vertex corrections (Fig.~\ref{quantum_effects}c), 
and other gluon loop corrections to the three-gluon and four-gluon vertices.
How the relevant amplitude is separated into different graphs is conventional. For instance, the 
quark self-energy, Fig.~\ref{quantum_effects}b, does not contribute in the Landau gauge.
Hence, stating which graphs contribute is partly arbitrary.
In fact, one can even arrange and combine graphs so that $\alpha_s$'s running is due only to gluon vacuum polarization, as in QED~\cite{Binosi:2016nme, Cui:2019dwv}.

\begin{figure}[t]
\centering
\includegraphics[width=15.0cm]{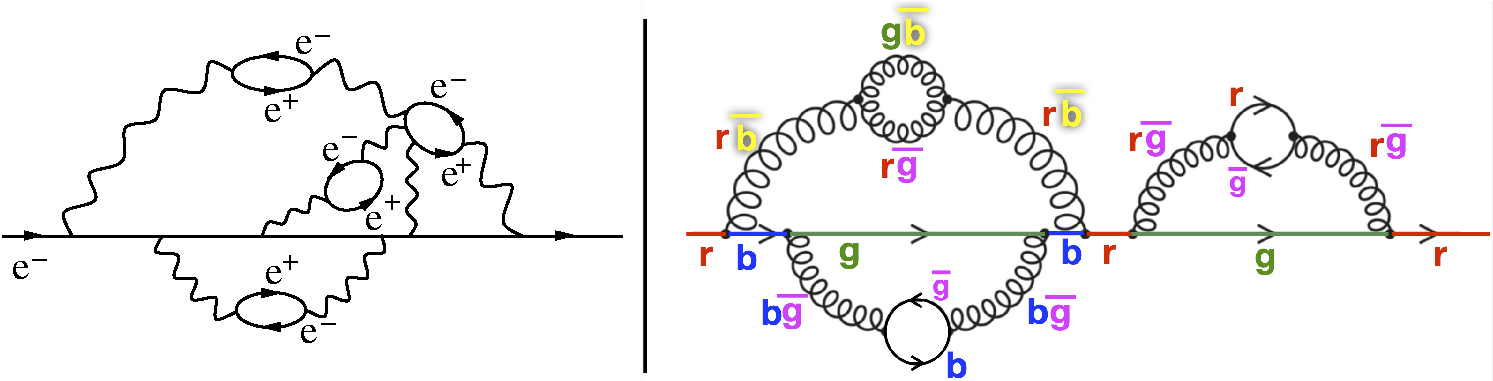}
\caption{\label{Flo:screening} 
\textbf{ Left: screening of the QED electric charge by quantum loops.} Positrons tend to be nearer to the bare negative charge (horizontal line) than electrons.
\textbf{Right: anti-screening of QCD color charges.} The color-charged gluons spatially spread the color initially carried by the quark: the initially red quark changes first into a blue quark when its red color is carried away by a gluon. Then, the quark becomes green after another gluon emission, etc. (Anti-blue is symbolized by yellow, and anti-green by magenta.) The spatial dilution of the initial red charge suppresses interaction occurring {\it via} high-resolution anti-red gluons, while lower-resolution gluons still couple since they do not resolve the quark from the gluons carrying away the color. Thus, $\alpha_s$ grows weaker at shorter distances.
(Figure from Ref.~\cite{Deur:2023dzc}.)
}
\end{figure}

How loops alter the $1/r^2$ law is sketched in Fig.~\ref{Flo:screening} (left: QED, right: QCD). 
For QED, pairs of virtual electron-positron temporarily appear around the bare charge, here that of an electron. 
The positrons are preferably closer to the bare charge because of their own opposite charge, while the virtual electrons tend to be farther. 
The total charge (bare+virtual) in a sphere of radius $r$ is thus smaller than the bare charge. By Gauss's law, the total charge controls 
the magnitude of the coupling, so a {\it charge screening} appears: 
$\alpha$ decreases as $r$ increases, asymptotically tending toward its long-range value $\sim$$1/137$.
For QCD, the charge-screening process is reversed because gluons carry colors. As depicted in Fig.~\ref{Flo:screening}, 
an emitted gluon carries away the initial color of the bare source, thereby spatially spreading the initial color charge. In Fig.~\ref{Flo:screening},
the bare quark is initially red but, eventually, is mostly green. Thus, it is transparent to high-resolution (relevant to large $Q^2$ processes) anti-red gluons, but
not to lower-resolution (low $Q^2$) anti-red gluons that 
would amalgamate the bare quark and the gluon(s) that carried away its red color. Hence, $\alpha_s$ decreases with $r$, opposite to QED. 
Screening from quark+antiquark loops occurs in QCD but the spacial spreading of color ({\it anti-screening}) from gluon loops dominates. 
Were more types of quarks to exist, {\it i.e.}, were the number of flavors $n_f$ larger in nature, 
screening would prevail once $n_f \geq 17$, and $\alpha_s$ would behave like 
QED's $\alpha$. Vertex correction (Fig.~\ref{quantum_effects}c) further enhances the color spatial spreading, 
while quark self-energy (Fig.~\ref{quantum_effects}b) either screens the charge or leaves it unaffected, depending on gauge choice.

Coupling of theories without intrinsic scales, like QED or QCD, runs logarithmically. Quarks, electrons, gluons and photons are pointlike and massless\footnote{Or nearly so for most of the fermions. The heavy quark masses can be considered infinite compared to the energy scale, and QCD remains without intrinsic scale.} so the momentum transfer in the reaction, $Q^2$, is the only available scale. Therefore, the infinitesimal scale dependence of a coupling $\textphnc{a}$ can only be: 
\begin{equation}
\frac{d\textphnc{a}(Q^2)}{dQ^2}= \frac{f(\textphnc{a})}{Q^2} = \frac{a_0+a_1 \textphnc{a} +a_2 \textphnc{a}^2 +\cdots \mathcal O(\textphnc{a}^n)}{Q^2}, 
\label{eq:coupling_expansion}
\end{equation}
where we assumed $\textphnc{a}$ to be small enough so that any function of it can be expanded. 
The (dimensionless) expansion coefficients $a_i$ are determined by the theory, see Section~\ref{alpha-s in the UV} for QCD. 
In fact, one can already infer that for QCD, $a_0 = a_1 = 0$, since for positive $a_0$ or $a_1$, $\alpha_s$ would increase at large $Q^2$, contradicting asymptotic freedom. 
And if $a_0$ or $a_1$ were negative, $\alpha_s$ would be unphysical (negative). 
Thus, for large $Q^2$ where $\alpha_s \ll 1$, $a_2$ dominates:
\begin{equation}
\alpha_s(Q^2) = \frac{-1}{a_2 \ln(\sfrac{Q^2}{C})},
\label{eq:simple_coupling}
\end{equation}
where $C$ is an integration constant and $a_2 < 0$ fulfills asymptotic freedom.
As $Q^2$ decreases, $\alpha_s$ increases and higher order $\ln^n(Q^2)$ terms become important. 
However, this description must ultimately break down when either 
({\bf A}) $\alpha_s$ becomes too large for Eq.~(\ref{eq:coupling_expansion}) to be valid, or 
({\bf B}) when $Q^{-1}$ reaches a value commensurate with the hadron size. 
Indeed, since confinement suppresses wavefunctions of colored particles when their wavelengths reach hadronic size, 
the very quantum effects responsible for the evolution of $\alpha_s$ are suppressed, making $\alpha_s$ constant 
there~\cite{Brodsky:2008be, Binosi:2014aea, Binosi:2016xxu, Binosi:2016wcx, Gao:2017uox}.
Clearly, conditions ({\bf A}) and ({\bf B}) are related since it is at the scale where $\alpha_s$ becomes large that the onset of confinement occurs, 
which in turn sets the hadron size. 

The techniques used to compute $\alpha_s$ in the short and long distance regimes are quite different. Therefore, we discuss them separately in Sections~\ref{alpha-s in the UV} and~\ref{alpha-s in the IR}. The large $Q^2$ (short-distance) regime is called the {\it ultraviolet} (UV) regime, and the low $Q^2$ (long-distance) one, the {\it infrared} (IR) regime. 
As of 2024, $\alpha_s$ is measured over 10 orders of magnitude in $Q^2$: $4\times 10^{-4}<Q^2/{\rm GeV}^2<4\times 10^{6}$~\cite{Deur:2022msf, Workman:2022ynf}.

A popular level account of $\alpha_s$ and its history is available in~\cite{Brodsky:2024zev}. 
To go deeper than the present chapter, several reviews on $\alpha_s$ are available; {\it e.g.},
Refs.~\cite{ParticleDataGroup:2024, Deur:2023dzc, Gross:2022hyw, dEnterria:2022hzv, Aoki:2021kgd, Proceedings:2019pra, Pich:2018lmu, Deur:2016tte, Dissertori:2015tfa, Altarelli:2013bpa, Prosperi:2006hx}. The standard review for $\alpha_s$ in the UV is from the Particle Data Group (PDG)~\cite{ParticleDataGroup:2024}. Most reviews cover
only the UV domain. Some also discussing the IR domain are~\cite{Deur:2023dzc, Gross:2022hyw, Deur:2016tte, Prosperi:2006hx}.

\subsection{Behavior of $\alpha_s$ at high-energy\label{alpha-s in the UV}}

The previous phenomenological description is formalized by the {\it renormalization} procedure
. In an interacting theory, the strength with which a field ({\it e.g.}, the force field) couples to another ({\it e.g.}, itself or a matter field) 
is set by the coupling constant. Its value becomes dependent on the {\it arbitrary} UV cut-off 
(or other methods) that regularizes the UV-divergent integrals. 
This unphysical feature is removed by making the coupling scale-dependent (to {\it run}) and by anchoring the running to a value 
phenomenologically determined 
at a chosen scale. The {\it coupling constant} becomes a running {\it effective coupling}.\footnote{
The procedure makes the coupling to loose its classical status of observable: it now depends on the chosen RS and, possibly, on the gauge choice, etc. Definitions that maintain observability exist, see {\it e.g.}, Section~\ref{EffectiveCharge}.} 
How the coupling runs is determined by requiring physics to be independent of human conventions, here the choice of renormalization scheme (RS). 
The symmetry group resulting from this invariance, the renormalization group (RG), allows us to compute the coupling behavior {\it via} group theory techniques. 
To see how, let us expand an observable~$R$:
\begin{equation}
R(Q^2)=\sum_{n}r_{n}\left[\alpha_s(Q^2)\right]^{n}. 
\label{eq:Exp_of_R}
\end{equation}
While $R$ is RS-independent,\footnote{In practice, Eq.~(\ref{eq:Exp_of_R}) is truncated to finite order so the perturbative approximant of $R$ has a residual RS-dependence.  \label{fn:residual-RS}} the series elements $r_n$ and $\alpha_s$ depend on the RS choice, except for $r_0$ and $r_1$ that are
RS-independent. For $r_0$, the reason is asymptotic freedom, $R \xrightarrow[Q^2 \to \infty]{} r_0$, and $r_0$ is an observable quantity, independent of RS.
For $r_1$, it is because at LO, $\alpha_s$ (Eq.~\eqref{Eq.one loop} below) is RS-independent since, 
although $\alpha_s$ may not be observable, its running arises from physical processes and can thus be expanded with another
coupling $\alpha^\prime_s$ obtained in a different RS:
\begin{equation}
\alpha'_s=\alpha_s+\mbox{v}_2\alpha^2_s+\mbox{v}_3\alpha^3_s+\cdots~.
\label{eq:alpha_rel_2RS}
\end{equation}
Then, using the other scheme, $R(Q^2)=\sum_{n}r'_n \left[\alpha^\prime_s(Q^2)\right]^{n}$ and Eq.~(\ref{eq:alpha_rel_2RS}) shows that 
$r_1$ is RS-independent.

QED's $\alpha$ slight RS-dependence makes it essentially constant in the IR~\cite{GellMann:1954fq}, allowing us to take it as an observable 
measurable in the IR. This is not so in pQCD: $\alpha_s$ is highly RS-dependent~\cite{Deur:2016cxb} and perhaps may not exist in the IR where 
quarks and gluons degrees of freedom are occulted. Therefore, $\alpha_s$ is often taken as an intermediate quantity having at best a 
qualitative physical meaning. For instance, $\alpha^{\rm MOM}_s(Q^2$=$1~\rm{GeV}^2) \simeq 1.5$ in the MOM 
RS~\cite{Celmaster:1979dm, Celmaster:1979km}, about 3 times larger than in the $\overline{\rm MS}$ RS~\cite{tHooft:1973mfk}, 
$\alpha^{\overline{\rm MS}}_s(1~\rm{GeV}^2) \simeq 0.5$. This shows that $\alpha_s$ cannot quantitatively stipulate the actual 
strength of QCD. 
Yet, asymptotic freedom implies that the RS-dependence vanishes in the deep UV regime (the $R \xrightarrow[Q^2 \to \infty]{} r_0$ above), and then, 
the physical and intuitive meaning of $\alpha_s$ is approximately restored there. 
Furthermore, we will see in Section~\ref{alpha-s in the IR} that $\alpha_s$ can be defined so that it retains its observable character and phenomenological meaning.

Considerations on interpretation/RS-dependence aside, the running of $\alpha_s$ is well understood in the UV thanks to pQCD.  
Yet, the latter provides only the $Q^2$-behavior, and experiments and lattice gauge theory (LGT) are deploying large efforts
to determine the absolute magnitude of $\alpha_s$. This must go hand-to-hand with advances in pQCD because $\alpha_s$ not being an observable,
high-order pQCD series are needed to accurately extract $\alpha_s$ from actual observables. 
The importance of such endeavors is clear considering that, as said earlier, $\alpha_s$ is by far the least known of the four fundamental couplings. 
Also, comparing  of $\alpha_s$ values obtained from distinct observables and at different $Q^2$ fundamentally tests QCD's internal 
consistency, and is thus a possible window into physics beyond the Standard Model
.  So, despite the well-known theoretical footing, $\alpha_s$ studies in the UV remain crucial. 
Let us now summarize the formalism providing the running of  $\alpha_s$.

The QCD Lagrangian density~\cite{Fritzsch:1973pi} is
\footnote{
We ignore gauge-fixing ghost fields since they are not fundamental nor required for our discussion.}:
\begin{equation}
\mathcal{L}_{\text{QCD}} = \sum_f \overline{\psi}_i^{(f)}\left(i\gamma_{\mu}D^{\mu}_{ij} - m_q\delta_{ij}\right)\psi_j^{(f)} - 
\frac{1}{4}F^{\mu\nu}_a F_{\mu\nu}^a,  \label{eq:QCD Lagrangian}
\end{equation}
where $\psi^{(f)}$ is the field for a quark of flavor $f$ and bare mass $m_q$, 
$D^\mu_{ij} \equiv \partial^\mu\delta_{ij} + i\sqrt{4\pi \overline{\alpha}_s}t^a_{ij} A_a^\mu$, with  
$\overline{\alpha}_s \equiv \overline{g}^2/4\pi$ the bare coupling \emph{constant},
$t^a_{ij}$ the SU(3) generators, $a=1,\ldots,8$ the color indices and $A_{\mu}^{a}$ the gluon fields, 
and 
$F^{\mu\nu}_a \equiv \partial^{\mu}A^{\nu}_{a}-\partial^{\nu}A^{\mu}_{a}+\sqrt{4\pi \overline{\alpha}_s}f_{abc}A^{\mu}_{b}A^{\nu}_{c}$, 
with $f_{abc}$ the SU(3) structure constants. 
We can generally take $m_q\simeq 0$ (light quarks) and $m_q \to \infty$ (heavy quarks) compared to $Q$ or the emerging 
QCD scale $\Lambda_s$ (Eq.~(\ref{eq:lambda-1loop})). 
Then, Eq.~(\ref{eq:QCD Lagrangian}) has no energy scale and defines a {\it conformally invariant} classical theory. 
Yet, phenomenologically, most QCD processes depend on $Q$, the process' 4-momentum flow. Therefore, a second scale must emerge 
at quantum level to normalize $Q$, thereby providing a dimensionless ratio. A ready candidate emerges from the renormalization procedure:
the {\it regularization scale} or {\it subtraction point} $\mu$. 
Its emergence in a classically conformal theory is called {\it dimensional transmutation}~\cite{Coleman:1973sx} and exemplifies a 
quantum anomaly
, {\it viz} the breaking by quantum effects of a symmetry in the classical theory, here conformal symmetry.  
In other words, QCD does not carry all the properties of its classical version embodied by $\mathcal{L}_{\text{QCD}}$. 
The meaning of $\mu$ varies with the choice of regularization method and RS~\cite{Deur:2023dzc} which, together 
with the fact that the value for $\mu$ is chosen arbitrarily, implies that any observable must be independent of $\mu$.
For example, consider a dimensionless observable $R$ that depends on the kinematic variables $Q^2$ and ${\bm y}$. The latter can be chosen to be 
dimensionless since they characterize a system without intrinsic physical scale. Expanding $R$ in $\overline\alpha_s$,
\begin{equation}
R(Q^2, {\bm y})=\sum_{n=0} r_{n}(Q^2/\mu^2, {\bm y})\overline\alpha_s^n,
\end{equation}
where the 
$r_n$ are calculated perturbatively. For $n\geq2$, divergences occur which, once regularized, depend on $Q^2/\mu^2$ 
(rather than $Q^2$, since there are no intrinsic scales). As explained in Section~\ref{running couplings}, the emerging scale dependence is assigned to the coupling $\overline\alpha_s$, apart from the classical $\propto 1/Q^2$ dependence. Thus, $\alpha_s$ acquires a running, and the pQCD series becomes:
\begin{equation}    
R(Q^2, {\bm y})=\sum_{n=0} r_{n}({\bm y, \alpha_s})\alpha_s^n(Q^2/\mu^2).
\end{equation}
The chain rule applied to the $\mu^2$-independent $R$ yields
\begin{equation}
 \frac{d R(Q^2, {\bm y})}{d\mu^2}=
 \left(\frac{\partial}{\partial \mu^2}+ 
 \frac{\partial \alpha_s}{\partial\mu^2} \frac{\partial}{\partial\alpha_s}+
 \frac{\partial Q^2}{\partial\mu^2} \frac{\partial}{\partial{Q^2}} +
 \sum_i \frac{\partial {y_i}}{\partial\mu^2} \frac{\partial}{\partial{y_i}}
  \right)R=0~.
 \label{eq:Callan-Symanzik1}
\end{equation}
Since $\bm y$ and $Q^2$ are physical quantities (kinematic variables), $\partial Q^2/\partial\mu^2 = 0 = \partial {y_i}/\partial\mu^2$. 
Then,  multiplying Eq.~(\ref{eq:Callan-Symanzik1}) by $\mu^2$ and using only the $Q^2/\mu^2$ ratio upon which $\alpha_s$ solely depends, 
yields the Renormalization Group Equation\footnote{
Also called the Callan--Symanzik relation, Gell-Mann--Low relation or 't Hooft-Weinberg relation, depending on the 
context and RS.} (RGE)~\cite{Petermann:1953wpa, GellMann:1954fq, Callan:1970yg, Symanzik:1970rt, Symanzik:1971vw}:
\begin{equation}
\left(-\frac{\partial}{\partial \big[\ln\left(Q^2/\mu^2\right)\big]}+\beta(\alpha_s)\frac{\partial}{\partial\alpha_s}\right)R=0, ~~~~~~~ \beta(\alpha_s) \equiv \mu^2\frac{\partial\alpha_s}{\partial\mu^2}, 
\label{eq:beta-func}
\end{equation}
which defines the ``$\beta$-function'' that governs the running of $\alpha_s$.\footnote{
We ignored the quark masses:  $m_q\simeq 0$ (light quarks) or $m_q \to \infty$ (heavy quarks). See~\cite{Deur:2023dzc} for a discussion on their second-order effect.}
Eqs.~(\ref{eq:Callan-Symanzik1}-\ref{eq:beta-func}) hold in any domain, both UV and IR. The $\beta$-function can be expanded in the UV as:
\begin{equation}
\beta\left(\alpha_s\right) = -\frac{\alpha_s^2}{4\pi}\sum_{n=0}\left(\frac{\alpha_s}{4\pi}\right)^{n}\beta_{n}.
\label{eq:alpha_s beta series}
\end{equation}
The 
$\beta_n$ are computed perturbatively and are presently available up to $n=3$ in several RSs, see~\cite{Deur:2016tte} for their expressions, 
and to $n=4$ in the $\overline{\rm MS}$ RS, see~\cite{Deur:2023dzc}. (We remind the reader that because the RS-dependence of $\alpha_s$ vanishes in the UV, 
$\beta_0$ and $\beta_1$ are RS-independent when the $m_q$ are ignored.) 
In the jargon, $\alpha_s$ calculated up to $\beta_n$ is said to be at the ``$n$+1-loop'' level since $n$+1 reflects the number of loops in the 
Feynman graphs of corresponding order.
By their definition in Eq.~\eqref{eq:alpha_s beta series}, the $\beta_n$ are independent of $\alpha_s$. Instead, they are expansions in $\hbar$. 
They depend solely on the number of active quark flavors, $n_f$, {\it i.e.}, quarks whose mass $m_q \ll \mu$  allows them to enter the loops making
 $\alpha_s$ to run. 
As we will compute below, $\beta(\alpha_s) < 0$ in the UV and for physical $n_f$ values. Then, from Eq.~(\ref{eq:beta-func}),
$\alpha_s \xrightarrow[\mu \to \infty]~$~0, {\it viz} asymptotic freedom. This finding~\cite{Gross:1973id, Politzer:1973fx} was pivotal for 
understanding the strong force: the vanishing of $\alpha_s$ makes perturbation theory applicable in the UV, 
which led to establishing QCD as the QFT of the strong force, thereby completing the Standard model of particle physics~\cite{Politzer2005}. 
This watershed in research on the strong force and particle physics was recognized by 
awarding the 2004 physics Nobel prize to the discoverers that in QCD\footnote{That $\beta(\alpha_s)$ can be negative has a rich history. Before the possibility percolated to QCD, Vanyashin and Terentyev~\cite{Vanyashin:1965ple}, and Khriplovich~\cite{Khriplovich:1969aa} discovered it for a SU(2) theory.  G. 't Hooft established it for an arbitrary gauge but reported his finding in his Ph.D. dissertation rather  than in a journal~\cite{tHooft:1998yft}.}, 
$\beta_0 > 0$, and thus $\beta\left(\alpha_s\right)<0$ in the UV~\cite{Gross:1973id, Politzer:1973fx}. 
In the jargon, $\alpha_s \xrightarrow[\mu \to \infty]~$~0 constitutes a {\it Gaussian fixed point}~\cite{Zinn-Justin:2010}. We will see in Section~\ref{alpha-s in the IR} that $\alpha_s$ also has a fixed point in the IR.

Let us now calculate the first-order (1-loop) $\beta$-function coefficient, $\beta_0$. This will expose what phenomena make $\alpha_s$ to run. 
We provide here the direct calculation, using Feynman rules with amplitudes already regularized and renormalized in $\overline{\rm MS}$, with $m_q = 0$. 
Another calculation method for $\beta_0$, pedagogically reported in~\cite{Deur:2023dzc} and summarized in Section~\ref{sec:DSE}, employs renormalization constants and is useful 
to consult since it makes explicit the connection between running and renormalization.
%
\begin{figure}[t]
\centering
\includegraphics[width=0.45\textwidth]{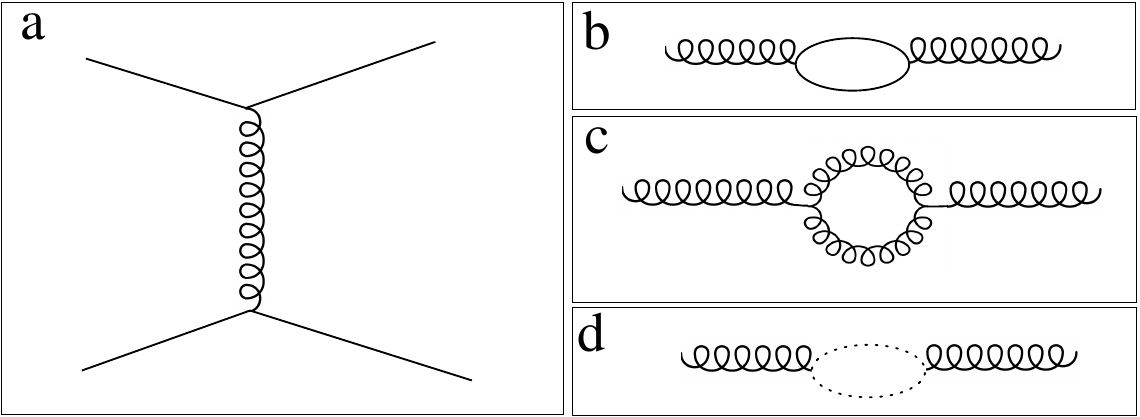}
\caption{\label{Flo:1st order corrections} \small Panel (a): 1$^{\rm st}$ order quark--quark scattering: classical result (no quantum loops) leading to the 
$1/r^2$ law.
Panels (b), (c) and (d): gluon propagator with a quark, gluon and ghost
loops, respectively. (Figure from Ref.~\cite{Deur:2023dzc}.)}
\end{figure}
%
The LO quark-quark interaction is drawn in Fig.~\ref{Flo:1st order corrections}a. As it has no loops, it is classical, yielding the $1/r^2$
law from the Fourier transform of $1/Q^2$ in the gluon propagator. Quantum effects arise from loops attached to the 
propagators or vertices of the LO diagram. As explained earlier, the extra $Q^2$-dependence they induce is folded into the 
coupling, making it run. 
The Feynman rules yield, for the gluon propagator with a quark loop (Fig.~\ref{Flo:1st order corrections}b):
\begin{equation}
D_q^{\mu\nu}\left(q\right) = -\frac{\overline{\alpha_s}}{3\pi}\left(q^{\mu}q^{\nu} - \eta^{\mu\nu}q^2\right) \ln(Q^2/\mu^2) n_f \frac{\delta_{ab}}{2},
\label{eq:quark loop}
\end{equation}
where $\eta^{\mu\nu}$ is the metric tensor, and $a$, $b$ are the in- and out-gluon color indices. The gluon loop (Fig.~\ref{Flo:1st order corrections}c)
adds a term:
\begin{equation}
D_{g}^{\mu\nu}\left(q\right) = \frac{\overline{\alpha_s}}{4\pi}N_c\delta_{ab}\left[\frac{11}{6}q^{\mu}q^{\nu} - \frac{19}{12}\eta^{\mu\nu}q^2
+ \frac{1-\xi}{2}\left(q^{\mu}q^{\nu} - \eta^{\mu\nu}q^2\right)\right] \ln(Q^2/\mu^2), \label{eq:gluon loop}
\end{equation}
with $\xi$ a gauge-fixing term and $N_c=3$, the number of colors. A longitudinal component is now present in the gluon propagator: 
$D_q^{\mu\nu}+D_{g}^{\mu\nu} \not\propto \left(q^{\mu}q^{\nu} - \eta^{\mu\nu}q^2\right)$, which violates current conservation:
$(D_q^{\mu\nu}+D_{g}^{\mu\nu})q_{\mu} \neq 0$.
There are several ways to fix this~\cite{Deur:2023dzc}. The most common is to introduce Faddeev-Popov ghosts~\cite{Faddeev:1967fc}, artificial\footnote{ 
Ghosts are spin-0 fields that nevertheless obey Fermi-Dirac statistics.}
particles designed so that their loop contribution (Fig.~\ref{Flo:1st order corrections}d),
\begin{equation}
D_{gh}^{\mu\nu}\left(q\right) = -\frac{\overline{\alpha_s}}{4\pi}N_c\delta_{ab}\left[\frac{1}{6}q^{\mu}q^{\nu} + \frac{1}{12}\eta^{\mu\nu}q^2\right] \ln(Q^2/\mu^2),
\label{eq:ghosts}
\end{equation}
cancels the longitudinal gluons, making the gluon propagator $(D_q^{\mu\nu} + D_{g}^{\mu\nu} + D_{gh}^{\mu\nu})$ purely transverse. 

In QCD, fermion self-energy (Fig.~\ref{quantum_effects}b) and vertex corrections (Fig.~\ref{quantum_effects}c) can affect the running of the coupling, 
contrary to QED where they cancel each other through the Ward identity~\cite{Ward:1950xp, Takahashi:1957xn}. 
In $\overline{\rm MS}$, self-energy and vertex corrections are, respectively:
\begin{equation}
G_q\left(p\right) = \frac{\not \! p}{p^2}\delta_{ab}\left[1-\xi \frac{\overline{\alpha_s}}{4\pi}\frac{N_c^2 - 1}{2N_c}\ln(-p^2/\mu^2)\right]~,
\label{eq:quark_self_E}
\end{equation}
%
%
\begin{equation}
\Gamma_{\mu}^{\alpha\beta;a}\left(q\right) = -i\sqrt{4\pi\overline{\alpha_s}}\frac{\lambda_{\alpha\beta}^{a}}{2}\gamma_{\mu}\left[1 - \frac{\overline{\alpha_s}}{4\pi}\ln(Q^2/\mu^2)\left\{ \xi\frac{N_c^2 - 1}{2N_c} + N_c\left(1 - \frac{1-\xi}{4}\right)\right\} \right],
\label{eq:vertex_cor}
\end{equation}
with $\lambda_{\alpha\beta}^{a}$ the Gell-Mann matrices and $\alpha, \beta$ the color indices of the in- and out-quarks. 
An advantage of $\overline{\rm MS}$ is that pQCD series coefficients, and so $\alpha_s$ too, are gauge-independent. 
Thus, $\xi$ can be assigned a convenient value, {\it e.g.}, $\xi=0$ (Landau gauge). Then,
the sum of the amplitudes from the propagators, Eqs.~\eqref{eq:quark loop}-\eqref{eq:quark_self_E}, and vertex, Eq.~\eqref{eq:vertex_cor}, 
provides the NLO corrected quark--quark interaction amplitude:
\begin{equation}
\mathcal{M} = \mathcal{M}_{Born}\left[1 + \frac{\overline{\alpha_s}}{4\pi}\left\{ \frac{2n_f}{3} - \frac{13N_c}{6} - \frac{3N_c}{2}\right\}\ln(Q^2/\mu^2) \right]~,
\label{Eq:NLO-qq}
\end{equation}
with $\mathcal{M}_{Born}$ the LO (classical) amplitude, Fig.~\ref{Flo:1st order corrections}a. 
The quark loops contribute the ``$2n_f/3$'' term, and the gluon and ghost loops contribute the ``$-13N_c/6$'' term, of opposite sign. 
The vertex correction term ``$-3N_c/2$'' also causes anti-screening. 
The quark self-energy correction does not contribute in $\overline{\rm MS}$, or any gauge-dependent RS when the Landau gauge is chosen, 
as evident from Eq.~\eqref{eq:quark_self_E}. The next step is crucial: the
quantum corrections in Eq.~(\ref{Eq:NLO-qq}) are folded into the constant $\overline{\alpha_s}$, thereby forming a running effective coupling,
\begin{equation}
\alpha_s\left(Q^2\right) = \alpha_s\left(\mu^2\right)\left[1+
\frac{\alpha_s\left(\mu^2\right)}{4\pi}\frac{2n_f-11N_c}{3}\ln(Q^2/\mu^2) \right],
\end{equation}
which, for $\alpha_s(\mu^2)\ln(\sfrac{Q^2}{\mu^2}) \ll 1$, yields:
\begin{equation}
\frac{4\pi}{\alpha_s\left(Q^2\right)} = \frac{4\pi}{\alpha_s\left(\mu^2\right)} + \frac{11N_c-2n_f}{3}\ln(Q^2/\mu^2).
\label{eq:alpha_s_1loop_1}
\end{equation}
After $Q^2$-differentiation,
\begin{equation}
-4\pi \frac{d\alpha_s\left(Q^2\right)}{\alpha_s^2\left(Q^2\right)} = \frac{11N_c-2n_f}{3}\frac{dQ^2}{Q^2}.
\end{equation}
Recalling the definition of $\beta$ and its expansion, Eq.~(\ref{eq:alpha_s beta series}), then:
\begin{equation}
 \boxed{\beta_0 = \frac{11N_c - 2n_f}{3}}~.\label{eq:beta0}
\end{equation}
Importantly, for the physical values of $N_c$ and $n_f$, $\beta_0 > 0$. The sign of $\beta$
determines how $\alpha_s$ runs. With $\beta_0$, the 1-loop contribution,  dominating in the UV, $\beta < 0$ and $\alpha_s$ decreases, leading to 
QCD's asymptotic freedom. Inspection~\cite{Deur:2023dzc} of the higher loop contributions shows that 
$\beta_1 > 0$ for $n_f \leq 8$, $\beta_2 > 0$ for $n_f \leq 5$,
and $\beta_3>0$ always, {\it viz} $\alpha_s$ still decreases at moderate
$Q^2$.

Solving Eq.~\eqref{eq:alpha_s_1loop_1} determines $\alpha_s$ at LO:
\begin{equation} 
\label{Eq.one loop}
 \boxed{\alpha_s(Q^2) = \frac{4\pi}{\beta_0 \ln\left(Q^2/ \Lambda_s^2\right)}}~.
\end{equation}
where 
\begin{equation}
\Lambda_s^2 \equiv \mu^2 \exp\left(-\frac{4\pi}{\beta_0\alpha_s(\mu^2)}\right).
\label{eq:lambda-1loop}
\end{equation}
Eq.~\eqref{eq:alpha_s beta series} is solved exactly and analytically only at 1-loop.
An exact solution at $\beta_1$-order (2-loop) is known~\cite{Gardi:1998qr}
but involves the Lambert function $W_{-1}$ with is therefore non-analytical. At $\beta_2$, or 3-loop, or higher orders,
no exact solutions are known and approximate solutions are obtained using an iterative method which
can be systematically applied to any order and currently provides solutions up to $\beta_4$ (5-loop) in $\overline{\rm MS}$~\cite{Kniehl:2006bg}.
The exact 2-loop solution and iterative higher-loop solutions 
are provided in~\cite{Deur:2016tte, Deur:2023dzc}. Finally, the definition of $\Lambda_s$, Eq.~(\ref{eq:lambda-1loop}), is for LO and differs at higher orders, 
see, {\it e.g.},~\cite{Aoki:2021kgd} for its exact nonperturbative expression. 
%
%
Comparing estimates of $\alpha_s$ up to $\beta_4$ (see Fig.~3.2 of Ref.~\cite{Deur:2023dzc}) reveals that starting at 2-loop, the $\beta$-series 
converges quickly in the UV: the effect of $n > 2$ loops and approximations is about 5\% at $Q^2=2$ GeV$^2$, near the edge of the UV domain. 
However, the difference between 1-loop and the higher-order results is typically above 20\% there.\footnote{
These numbers reflect the effects of loops, not an uncertainty in $\alpha_s$, which is significantly smaller (below 1\%).
 The truncation uncertainty is obtained by computing the N$^n$LO pQCD series of an observable using $\alpha_s$ at $n$-loop, and comparing it to 
 the result at N$^{n-1}$LO and ($n-1$)-loop. It provides the overall truncation uncertainty from both the pQCD series ($\alpha_s$ expansion) and the $\beta$ series ($\hbar$ expansion).}

Once Eq.~\eqref{eq:alpha_s beta series} is solved at the desired order and with the needed $\beta_i$ calculated, $\Lambda_s$ 
remains the only unknown quantity. According to QCD and as verified experimentally, $\alpha_s(Q^2)$ in the UV is monotonic with $Q^2$, 
so knowing either $\Lambda_s$ or $\alpha_s(Q^2_{\rm ref})$ at a given $Q^2_{\rm ref}$ value, usually chosen as $Q^2_{\rm ref} = M_Z^2$, 
provides $\alpha_s(Q^2)$ anywhere on the UV domain. Therefore, a value for $\Lambda_s$ can be provided in lieu of the absolute magnitude 
$\alpha_s(Q^2_{\rm ref})$. Both are phenomenological quantities to be determined experimentally.
Alternatively, nonperturbative theoretical methods, {\it e.g.}, LGT, can deduce them from other accurately known phenomenological inputs, {\it e.g.}, the nucleon mass.
In other words, $\Lambda_s$ (or $\alpha_s(Q^2_{\rm ref})$) is neither determined nor explained within QCD, even nonperturbatively, 
just as $\alpha\simeq1/137$ is not explainable within QED.

{\bf The Landau p\^ole \label{Landau pole}}
$\Lambda_s$ determines the rate at which $\alpha_s(Q^2)$ changes with $Q^2$. It also provides 
the scale where $\alpha_s^{\rm pQCD} \hspace{-1.5mm}\to \infty$, where pQCD has clearly failed. This, however, is a qualitative rather than a fully 
objective indication: like $\alpha_s$, $\Lambda_s$ is typically not an observable. Beyond 2-loops, it is RS-dependent. It also depends 
on the loop order at which $\alpha_s$ is expanded and the method used to solve Eq.~\eqref{eq:alpha_s beta series}. 
Values for $\Lambda_s$ in different RS typically range from 0.3 to 1 GeV; see Table 3.1 in Ref.~\cite{Deur:2023dzc}, 
which also provides the all-order relation between $\Lambda_s$ in different RS.
The place where the pQCD expression of $\alpha_s$ diverges, exactly at $Q^2=\Lambda_s^2$ at 1-loop (see Eq.~\eqref{Eq.one loop}) and 
nearby for higher-order approximations, is called the Landau p\^ole~\cite{Landau:1959fi, Landau:1965nlt}. It was first encountered in QED, at  
$\Lambda_{\rm EM}\sim10^{30-40}$ GeV, well above the Planck scale and thus well within a domain 
where the Standard Model, and perhaps QFT, should be superseded by a more fundamental theory. 
Thus, the QED Landau p\^ole is irrelevant. Since $\alpha_s$ runs opposite to $\alpha$, the QCD Landau p\^ole sits at low energy 
($\Lambda_s\approx0.5$~GeV) but is also irrelevant. The p\^ole just signals where the QFT perturbative treatment has failed since 
an observable's series expanded in $\alpha_s$ would ``doubly'' diverge because 
{\bf (A)} each next order correction would typically grow larger, and 
{\bf (B)} $\alpha_s \to \infty$. Such a divergence is not observed.
\footnote{Hadronic observables, {\it e.g.}, form factors and structure functions, have been measured over $Q^2$ domains that comprise $\Lambda_s$ without showing any unusual behaviors such as divergences or discontinuities. The Landau p\^ole unphysical nature is evinced by other facts: 
the value of $\Lambda_s$ depends on the (arbitrary) RS choice; and were the p\^ole genuine, it would produce 
particles of imaginary mass $m=q\equiv \sqrt{-Q^2}=i\Lambda_s$,  {\it i.e.},  tachyons.
} 
Likewise, 
the $\beta$-function series is similarly affected (but not its $\beta_n$ coefficients, which are expanded in $\hbar$). 
Despite the Landau p\^ole being unphysical, $\Lambda_s$ is often called {\it confinement scale} because it lies within the nonperturbative regime where confinement --a nonperturbative phenomenon-- occurs and because it suggests that there, $\alpha_s$ has become large enough to trigger the confinement process. 
The Landau p\^ole has sometimes been incorrectly considered physically relevant. For instance, the QED Landau p\^ole challenged the adequacy of 
QFT\footnote{
The p\^ole famously steered L. Landau away from QFT and, through his ascendency, most of Soviet research on particle physics.}
as a description of nature~\cite{Leutwyler:2012ax}. As already mentioned, this worry vanishes once one realizes that QED 
is a low-energy effective theory, to be superseded  at energies well below its Landau p\^ole. 
Another example is the Landau p\^ole being sometimes thought, in QCD's early days, as causing quark confinement. 
This was dubbed ``IR slavery,'' a pleasing but erroneous counterpart of UV's asymptotic freedom~\cite{Marciano:1977su}. 
Nevertheless, within a specific theoretical framework, one may relate  $\Lambda_s$ to a physical scale, {\it e.g.}, that characterizing the 
hadron mass spectrum. This has been shown  {\it via} lattice gauge theory (LGT)~\cite{Aoki:2021kgd}, the Dyson-Schwinger equations (DSE)~\cite{Cui:2019dwv}, 
or holographic light-front QCD (HLFQCD)~\cite{Deur:2014qfa}. In the latter approach, the Landau p\^ole evolves from a real $Q^2>0$ p\^ole  
(therefore unphysical) to an imaginary p\^ole $i\Lambda_s$ in the complex $Q^2$-plane, as the scale for hadron masses and confinement evolves from zero to its physical value~\cite{deTeramond:2024ikl}, see Section~\ref{HLFQCD} for details.

\subsection{Methods of determination of $\alpha_s$ at short distances}
As mentioned, the absolute scale of $\alpha_s$, or equivalently $\Lambda_s$, is not calculable within QCD. 
They are therefore obtained either experimentally or by relating them using nonperturbative techniques  to other phenomenologically determined scales, like the nucleon mass. Observables providing $\alpha_s$ are numerous. 
Indeed, any hadronic observable with a pQCD expansion is usable. In practice, however, some observables are better suited for precision extractions 
of $\alpha_s$. On the theory side, LGT is preferred as it is a well-controlled approximation to QCD unlike many models of hadron structure 
whose uncertainties are difficult to assess. Yet, some models have provided compelling determinations~\cite{Brodsky:2014yha, Cui:2019dwv}. 
Here, we briefly outline the methods to obtain $\alpha_s$, giving examples of the most common observables that yield a precise $\alpha_s$. 
More exhaustive lists are in the PDG review and compilation of $\alpha_s$ in the UV~\cite{Workman:2022ynf} and in the LGT-oriented review from the FLAG collaboration~\cite{Aoki:2021kgd}. In both cases, the compiled results are in the $\overline {\rm MS}$ RS (which reminds us that $\alpha_s$ is generally not an observable).
\begin{figure}[t]
\centering
\includegraphics[width=0.95\textwidth]{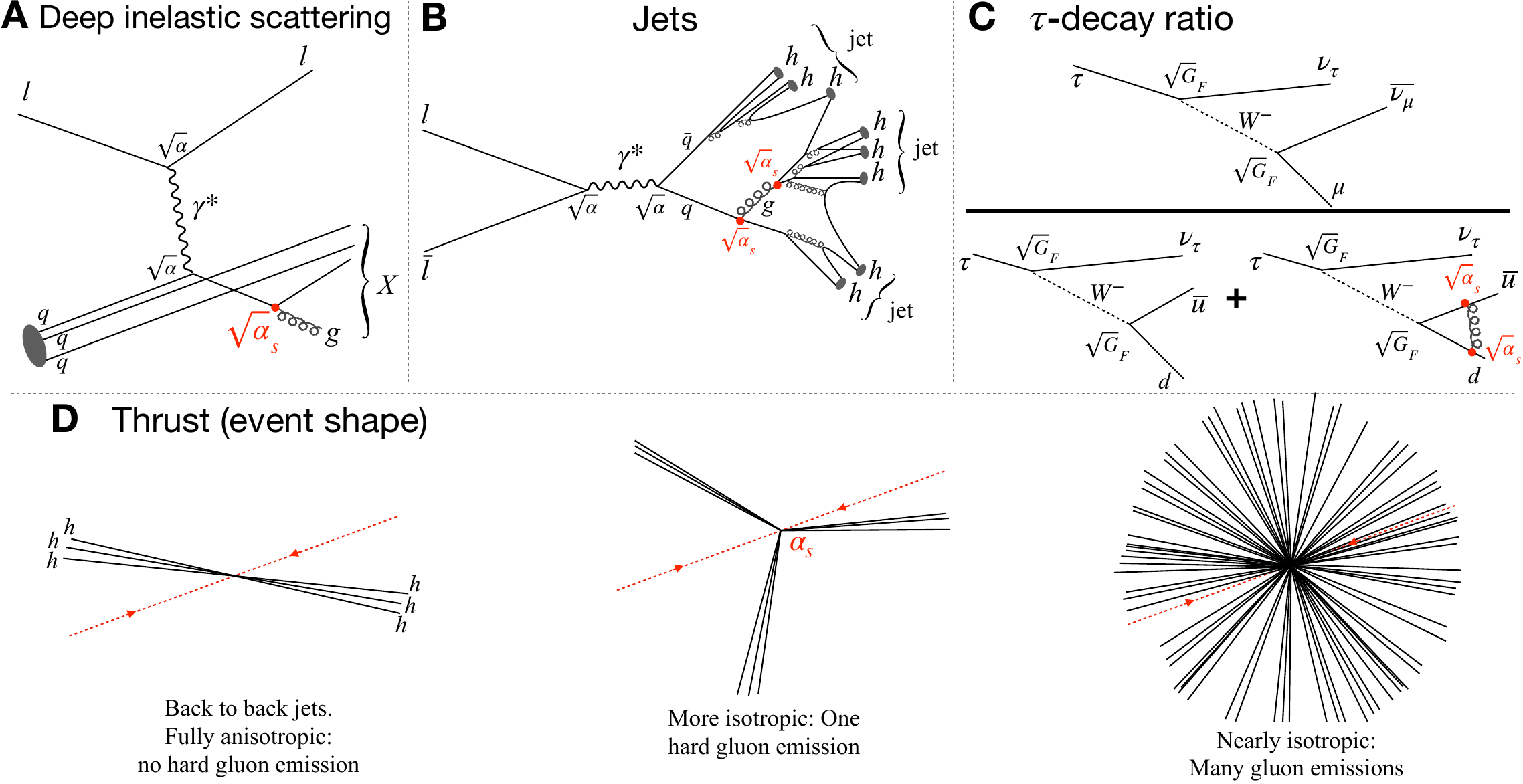}
\caption{\label{Flo:exp_reac} \small
{\bf High-energy reactions typically used to determine $\bm \alpha_s$.} A curly line with a $g$ label denotes the hard gluon that allows access to $\alpha_s$. (``Hard'' means a high-energy gluon emitted at large angle from its parent quark.) Softer gluons are shown by  unlabelled curly lines. 
{\bf A}: Deep inelastic scattering of a charged lepton $l$ by a nucleon (grey blob). $X$ denotes the undetected final hadronic state.
{\bf B}: Lepton-antilepton annihilation, eventually forming hadronic jets ($h$ denotes a hadron). 
{\bf C}: Comparison between leptonic and hadronic $\tau$-decays.
{\bf D}: Geometrical variation of the jet distributions. The red dashed lines are the two colliding particles and the solid lines the produced hadrons. 
}
\end{figure}

For $Q^2 \gg \Lambda_s^2$, the influence of $\Lambda_s$ -- and {\it a fortiori} that of light quark masses -- must vanish. 
Furthermore, quarks being pointlike and asymptotically free, no structures exist at large $Q^2$, and hence, no new scales arise. 
The appearance of a conformal 
behavior, known as {\it Bjorken scaling}~\cite{Bjorken:1969ja, Feynman:1969wa}, 
implies a $Q^2$-independence of hadronic structure quantities. Scaling violations at large $Q^2$ reflect the residual influence of $\Lambda_s$ 
(and to a lesser extent, that of hadron masses and structures). This offers a first way to access $\Lambda_s$ or $\alpha_s$ typified 
by charged lepton--quark scattering, Fig.~\ref{Flo:exp_reac}A. 
The reaction is elastic, quarks being structureless, and occurs during deep inelastic scattering (DIS) of a lepton off a nucleon.
At LO, scaling violations originate from the struck quark emitting a gluon, pair creation from that gluon, and photon--gluon 
``fusion'' (photon--gluon interaction {\it via} quark pair creation). At NLO, quark self-energy and quark--photon vertex corrections  also contribute. 
The ensuing $Q^2$-dependence obeys the Dokshitzer--Gribov--Lipatov--Altarelli--Parisi (DGLAP) equations~\cite{Dokshitzer:1977sg, Gribov:1971zn, Lipatov:1974qm, Altarelli:1977zs}, which provide the formalism to extract $\alpha_s$. 
In practice, the $Q^2$-dependence of nucleon structure quantities is fit using DGLAP, together with parameterizations of the (nonperturbative) 
momentum distributions of the quarks and gluons in the nucleon (PDFs: parton distribution functions).
Several collaborations use this extraction method, see~\cite{Deur:2023dzc, Workman:2022ynf}.

Observing hadronic jets is another way to access $\alpha_s$ {\it via} hard gluon emissions. For instance, Fig.~\ref{Flo:exp_reac}B shows a 3-jet event 
where one jet developed from a hard gluon. The gluon emission probability, and thus also of $\gamma^\ast q \to q g$, being proportional to $\alpha_s$, 
the 3-jet event rate normalized to the 2-jet rate directly provides $\alpha_s$. 
The geometry of the particle collision outcomes can also provide $\alpha_s$. Such geometry is analyzed with {\it event shape} 
observables, {\it e.g.}, the {\it thrust}, which quantifies the anisotropy of the particle emission produced in the collision, 
Fig.~\ref{Flo:exp_reac}D. In the infinite momentum-transfer limit, all particles are aligned along a preferred axis, 
revealing the back-to-back produced $q\bar{q}$. 
At finite momentum-transfer, gluon emission from the quarks isotropizes the jet distribution geometry, with a spherical symmetry at low momentum-transfer. 
The deviation from anisotropy being due to gluon emissions, it allows access to $\alpha_s$. Other event shape observables are used similarly. 
Also, inclusive cross-sections of heavy quark (bottom, top) production in collisions now accurately provide  $\alpha_s$, 
owing to refinements in theoretical understanding~\cite{Deur:2023dzc}.

Other ways to access $\alpha_s$ are to compare the hadron production rate in $l \bar l$ annihilation to that of muons, 
or comparing hadronic to leptonic decay widths of a particle ($Z^0$, $W^\pm$,  $\gamma^* \cdots$) produced in the annihilation~\cite{Pich:2020gzz}. 
The sensitivity to $\alpha_s$ arises mostly through hard gluon emission by a quark or antiquark. The example of the ratio of $\tau$-decay into hadrons 
over that of $\tau \to  \nu_\tau e^-\bar{\nu_e}$ is sketched in Fig.~\ref{Flo:exp_reac}C. It provides the most precise experimental extraction of $\alpha_s(M_Z)$
because evolving $\alpha_s(M_\tau)$ to $\alpha_s(M_Z)$ reduces its uncertainty by an order of magnitude.\footnote{
Specifically, $\delta \alpha_s(M_Z)= \left(\frac{\alpha^2_s(M_Z)}{\alpha^2_s(M_\tau)}\right) \delta \alpha_s(M_\tau)\simeq 0.11  \delta \alpha_s(M_\tau)$ 
because $\left| \frac{d(1/\ln(Q^2/\Lambda_s^2))}{dQ^2} \right| = \frac{1}{\ln^2(Q^2/\Lambda_s^2)}$.
}
However, its rather low $Q^2=M_\tau^2=3.16$~GeV$^2$ value raises accuracy issues that remain debated~\cite{Deur:2023dzc, Pich:2020gzz}. 
It causes non-trivial higher-order pQCD and nonperturbative QCD contributions. Furthermore, the produced $\nu_\tau$ allows the $W^-$ momentum to take any  
kinematically allowed values, including zero. Yet, a pQCD treatment is possible thanks to analyticity arguments. 
Several perturbative schemes are used for such analyses but the resulting $\alpha_s$ values are generally in tension. 
Which scheme is accurate and how to resolve the tension remain unclear~\cite{Bethke:2011tr, Hoang:2020mkw, Boito:2020hvu, Pich:2016bdg}. 
In global averages of $\alpha_s$ from $\tau$-decay, the difference between the various schemes is split and taken as an uncertainty.

Three types of uncertainties contribute in the experimental extraction of $\alpha_s$: 
({\bf A}) experimental uncertainties;
({\bf B}) truncation of the pQCD series to a finite order; and 
({\bf C}) possible nonperturbative contributions. 
Observables competitive in extracting $\alpha_s$ balance these contributions, often differently. The list also makes clear that improvement experimental 
precision data must be accompanied by theoretical advances in determining the observables' pQCD series 
at sufficiently high order, and efforts to control the nonperturbative corrections. 

Presently, the most accurate way to obtain $\alpha_s$ is LGT, 
with a precision of $\sim$0.7\%~\cite{Aoki:2021kgd}. 
Ref.~\cite{Deur:2016tte} provides an introduction to LGT in the context of determining $\alpha_s$. The the FLAG collaboration~\cite{Aoki:2021kgd} compilation 
is the authority for $\alpha_s(M_Z)$ from LGT, and provides a comprehensive summary of the LGT methods. 
%
LGT accesses $\alpha_s$ like experiments do: an appropriate quantity\footnote{For LGT, the quantity need not be an observable, unlike for experiments.} 
is computed and matched to the corresponding pQCD series. Optimizing the match yields $\alpha_s$. 
It results from the procedure that although LGT is a nonperturbative formalism, its $\alpha_s$ is RS-dependent, with the RS that of the pQCD series. 
As with experimental extractions, different quantities offer different advantages, and groups performing LGT calculations have opted for different quantities. 
Currently, the best ones for $\alpha_s(M_Z)$ are Wilson loops, heavy-quark potential at short distances, heavy-quark current two-point functions, 
and $\alpha_s$ itself in association with the step-scaling method~\cite{Luscher:1991wu}. Other quantities are various QCD vertices, 
the Dirac operator, and vacuum polarization, but they are for now not as precise. 

LGT is not the only nonperturbative method available for $\alpha_s(M_Z)$ computation, but it is currently the only one trusted by the PDG in 
its compilation~\cite{Workman:2022ynf}. Other methods providing precise determinations of $\alpha_s(M_Z)$ include 
the Dyson-Schwinger equations (DSE~\cite{Roberts:1994dr, Alkofer:2000wg}) 
combined with LGT~\cite{Zafeiropoulos:2019flq}, and 
AdS/QCD~\cite{Brodsky:2016yod, Deur:2014qfa, Deur:2016opc}. 

\begin{figure}[t]
\centering
\includegraphics[width=0.35\textwidth,height=0.3\textwidth]{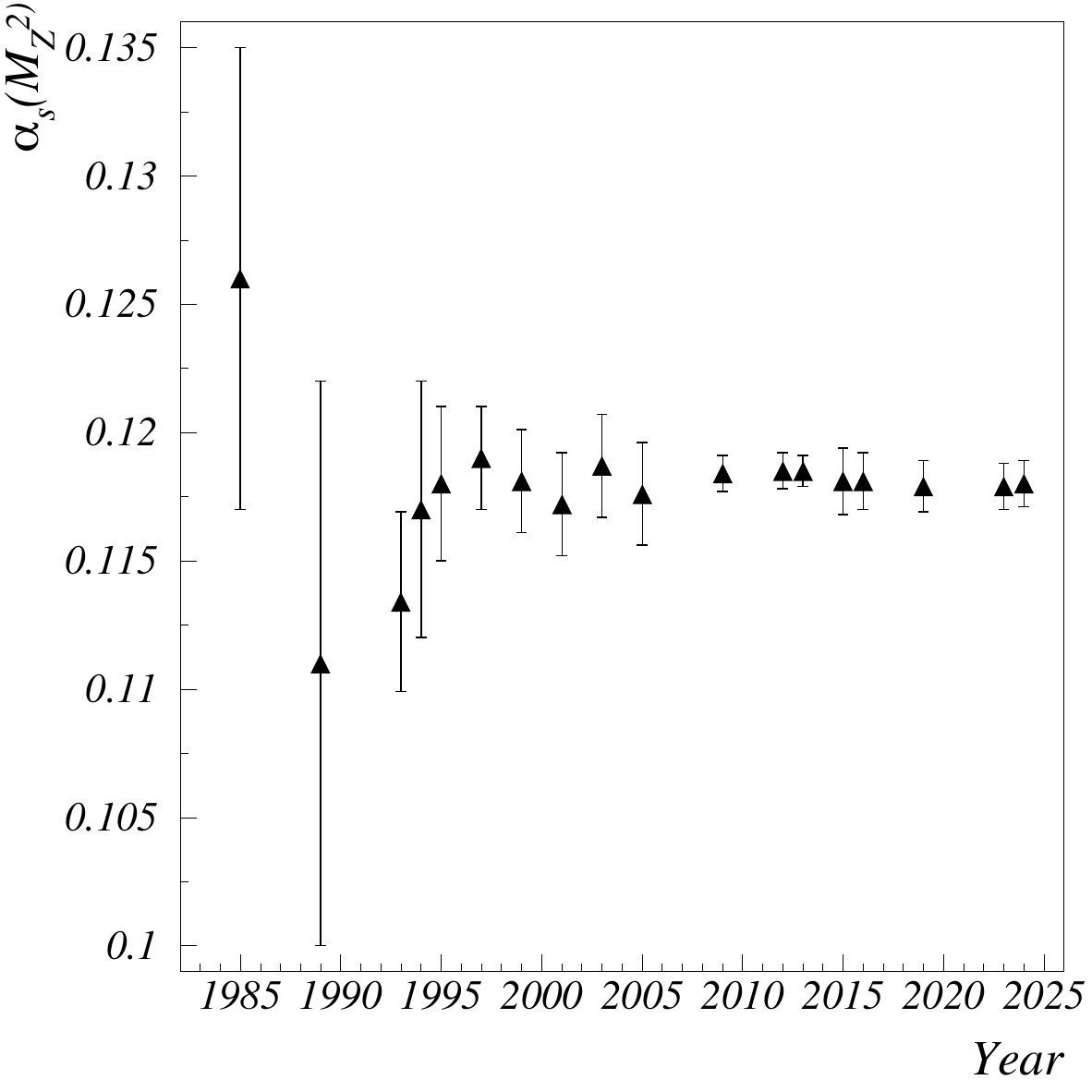}
\caption{\label{Flo:history} \small
{\bf History of the global average of $\bm \alpha_s(M_Z)$} (data from the Particle Data Group).
The notable improvement after 1989 reflects the impact of CERN's LEP ($e^+ e^-$ collisions) and 
DESY's HERA ($e^{+/-} p$ collisions) starting operations, together with theoretical advances on NNLO calculations. 
Next, in the mid-2000s, LGT determinations came to dominate the global average, albeit in 2015, the uncertainty was revised upward as it was 
assessed that the LGT systematics were larger than initially estimated. Then, the steady decrease in uncertainty resumed, with presently $\alpha_s(M_Z) = 0.1180(9)$.}
\end{figure}

We conclude this section with the historical progress and perspectives in determining $\alpha_s(M_Z)$. Past and present PDG global averages 
are shown in Fig.~\ref{Flo:history}. 
Currently,  
\fbox{\parbox{4.3cm}{$\bm {\alpha_s(M_Z) = 0.1180 \pm 0.0009}$}}
(\mbox{$\overline{\rm MS}$~RS})~\cite{Workman:2022ynf}. Accuracy will continue to improve as new data become available, as new facilities, 
{\it e.g.}, EIC~\cite{Accardi:2012qut, Kutz:2024eaq} or EicC~\cite{Chen:2020ijn, Anderle:2021wcy} come online or existing ones are upgraded, 
{\it e.g.}, LHC~\cite{EuropeanStrategyforParticlePhysicsPreparatoryGroup:2019qin} or JLab~\cite{Accardi:2023chb}. 
As mentioned, for experimental extractions of $\alpha_s(M_Z)$ to reach their full potential, pQCD theoretical developments must go hand-in-hand.


\subsection{Long distance behavior of \mbox{$\mathbf \alpha_s$}}
\label{alpha-s in the IR}
\begin{figure}[t]
\centering
\includegraphics[width=0.35\textwidth,height=0.3\textwidth]{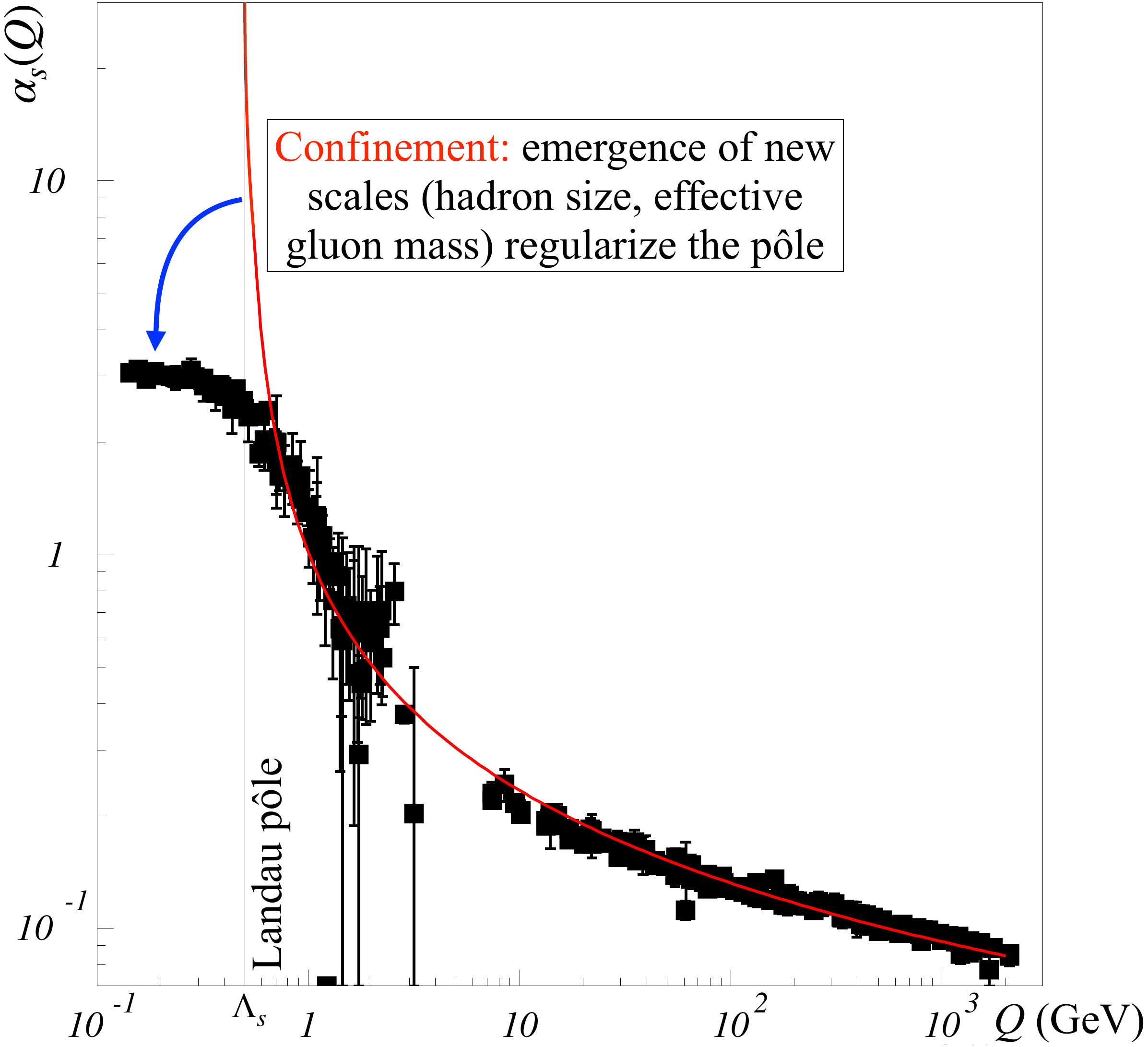}
\caption{\label{Flo:pole_regul} \small
{\bf Regularization of the Landau p\^ole.}}
\end{figure}
So far we have discussed the UV domain of weakly-coupled QCD where $\alpha_s$ is most familiar thanks to pQCD's power. 
Now, what happens outside the UV? pQCD ``predicts'' that $\alpha_s$ diverges when $Q^2 \to  \Lambda_s^2$, see Section~\ref{Landau pole}. 
This divergence is generic to couplings calculated perturbatively: we saw that it also occurs in QED. 
It merely signals the applicability limit of perturbation theory. Hence, color confinement cannot come from the 
Landau p\^ole which would, anyway, generate unphysical tachyons. As already mentioned, no Landau p\^ole is expected in reality  
for QED since it will be superseded by a more fundamental theory, nor for QCD because the $Q^2$-dependence of the actual coupling is suppressed 
in the IR by the physical hadron size. This one imposes a maximum wavelength in the loops causing the running, 
thereby stopping it~\cite{Brodsky:2008be, Binosi:2014aea, Gao:2017uox}, Fig.~\ref{Flo:pole_regul}. 
What happens to the p\^ole is discussed in Section~\ref{HLFQCD}. The absence of $Q^2$-dependence of $\alpha_s$ in the IR, {\it i.e.}, the vanishing QCD 
$\beta$-function, is variously named the {\it freezing of $\alpha_s$}, 
the {\it conformal window} of QCD, or the {\it $Q^2=0$ fixed point}.

Nonperturbative studies of $\alpha_s$ are more arduous than those using pQCD. Yet, they are crucial because in nature, the strong force manifests 
 itself mostly nonperturbatively. For example, almost all of the visible mass in the universe emerges from QCD IR dynamics\footnote{
The Higgs mechanism contributes at the few \% level.}~\cite{Roberts:2021nhw}, with unsurprisingly the IR behavior of $\alpha_s$ 
being crucial to the process~\cite{Deur:2014qfa, Binosi:2014aea, Binosi:2016xxu}. 
Playing an important role in this phenomenon is dynamical chiral symmetry breaking, whose realization hinges on the magnitude of $\alpha_s$ in the IR~\cite{Nambu:1961tp, Lane:1974he, Politzer:1976tv, Binosi:2016wcx}. 
The chief reason why IR studies of $\alpha_s$ are difficult is its various possible definitions, 
without one seemingly superior to the others. This contrasts with pQCD with its 
single agreed definition. The issue is largely due to having no definite nonperturbative solution of QCD, with many methods being tried. 
Should one method lead to a clear analytical nonperturbative solution of QCD, it would yield a compelling candidate for $\alpha_s$. 
The picture just drawn may seem bleak but in an encouraging recent development, several nonperturbative methods, namely, 
LGT, DSE, and AdS/CFT, have produced IR couplings that are consistent and also agree with IR experimental data on $\alpha_s$. 
This offers a compelling case for having finally identified a canonical IR definition 
for $\alpha_s$. In this Section, we discuss the origin of the aforementioned challenges and then describe the recent developments.

\begin{figure}[t!]
\centering
\includegraphics[width=0.4\textwidth, height=0.34\textwidth]{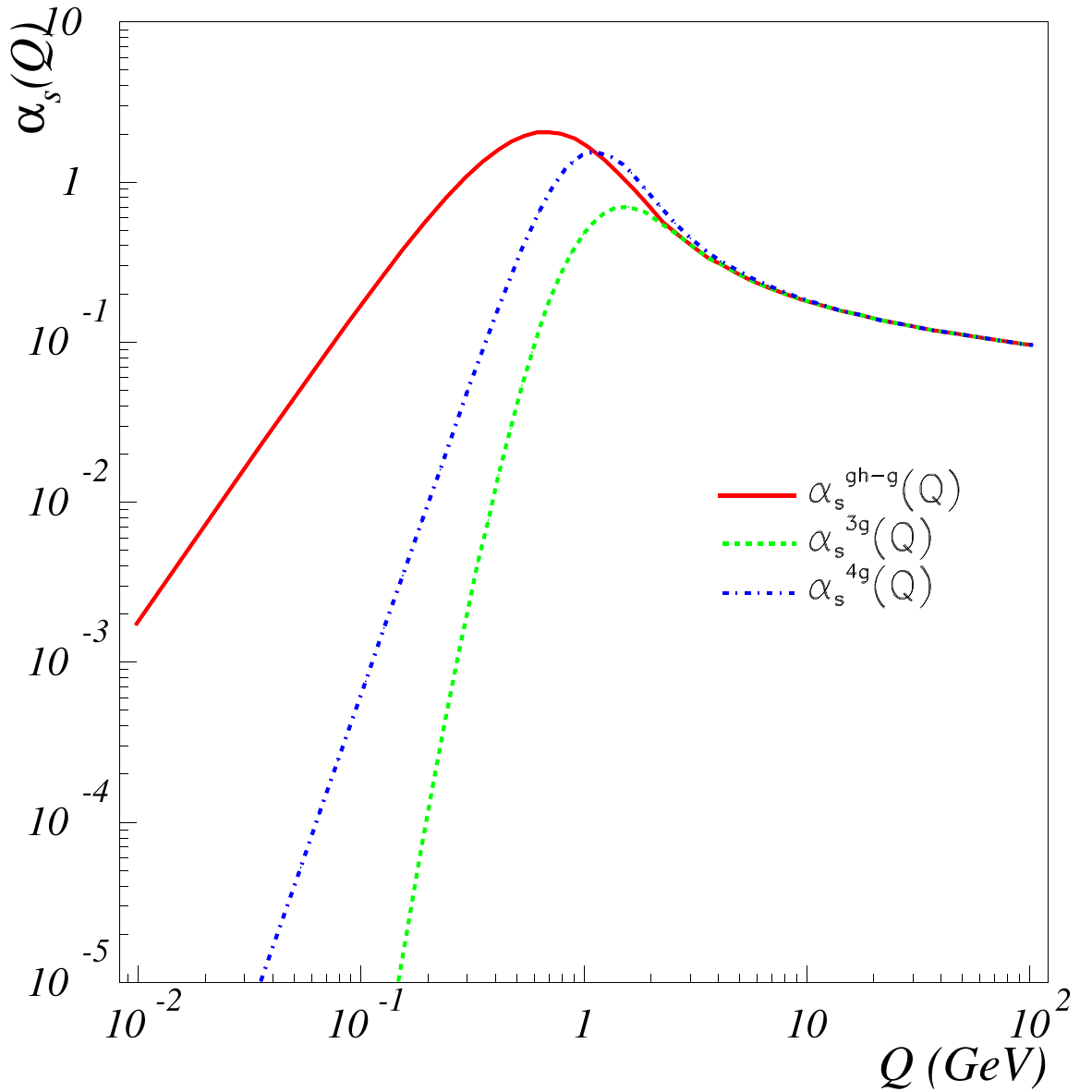}
\includegraphics[width=0.4\textwidth, height=0.34\textwidth]{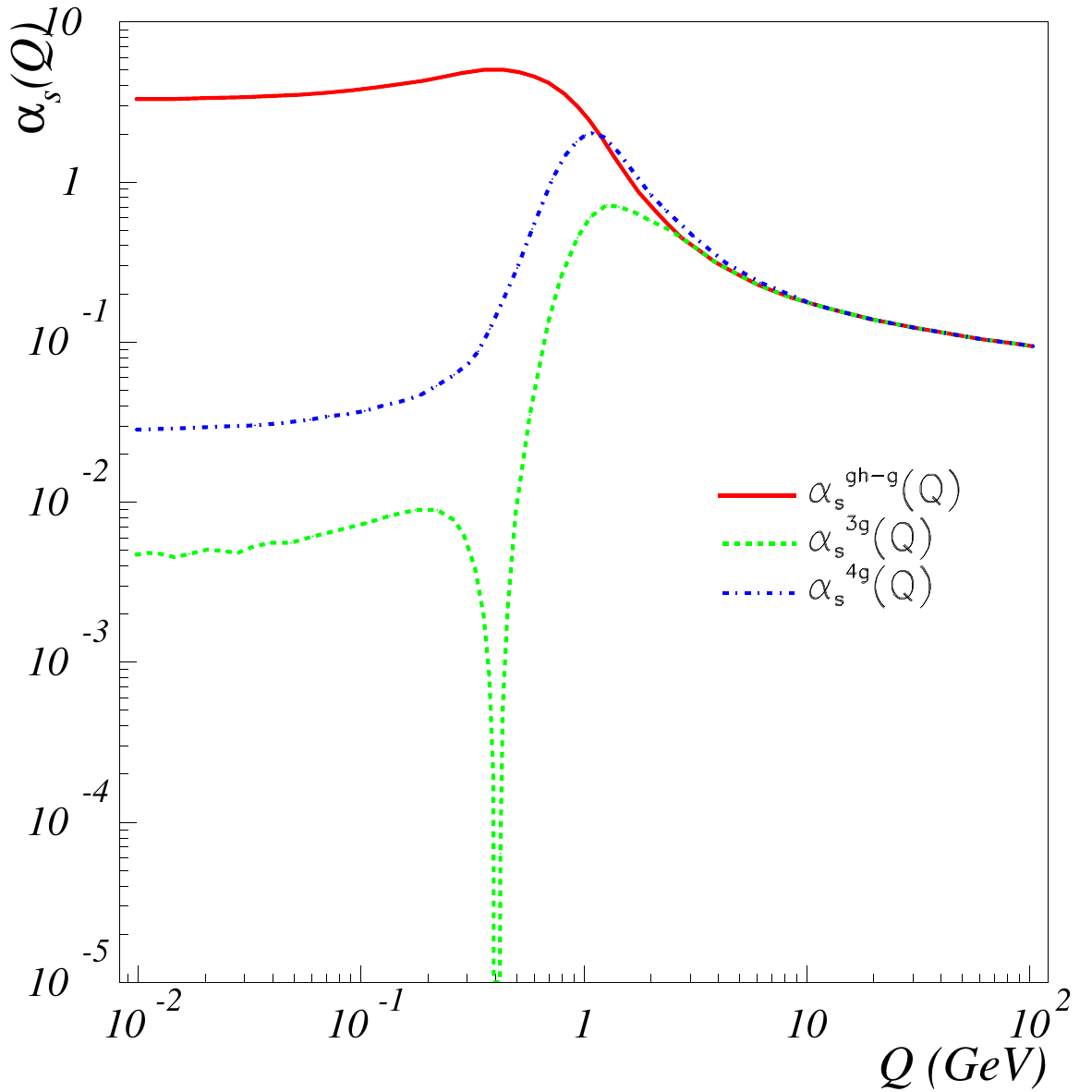}
\caption{\textbf{Different vertex couplings} 
Decoupling and scaling solutions are shown in the left and right panels, respectively~\cite{Huber:2020keu}.  In both cases, the ghost-gluon ($\alpha_s^{gh}$), 3-gluon ($\alpha_s^{3g}$), and 4-gluon  ($\alpha_s^{4g}$) couplings  
behave very differently, except in the UV where the STI always hold. 
Calculations are in Landau gauge and a MOM RS. 
}
\label{Fig:various_vertex_couplings}
\end{figure}

As just said, studying $\alpha_s(Q^2)$ in the IR is challenging  chiefly because no obvious 
definition of $\alpha_s$ is available. (Remember, $\alpha_s$ need not be an observable, see Section~\ref{alpha-s in the UV}). 
Other reasons are 
({\bf A}) systematics or model bias in nonperturbative calculations are often hard to control; 
({\bf B}) the RS-dependence increases at lower $Q^2$; 
and ({\bf C}) the various vertices (3-gluon, 4-gluon, quark-gluon, or ghost-gluon) may have different couplings, see Fig.~\ref{Fig:various_vertex_couplings}.
That is, distinct magnitudes and $Q^2$-dependence may characterize these vertices.\footnote{
This may seem odd, but it just means that the various quantum loops contribute differently to different vertices which, at classical 
level (bare coupling $\overline{\alpha}_s$) or observational level (deep UV limit, or defining $\alpha_s$ as an observable) do couple with a universal strength. }
This occurs when the chosen gauge and RS  conculcate the Slavnov-Taylor identities (STI)~\cite{Taylor:1971ff, Slavnov:1972fg}, 
QCD's version of QED's Ward identity~\cite{Ward:1950xp, Takahashi:1957xn}. In the $\overline{\rm{MS}}$ RS, the STI hold, 
so all vertex couplings are the same. However, $\overline{\rm{MS}}$ is not suited to most nonperturbative methods. 
Instead, the MOM RS and Landau gauge are often used, in which case, the various vertex couplings may behave differently (Fig.~\ref{Fig:various_vertex_couplings}). 
%
%
Historically, many IR definitions of $\alpha_s$ have been used, producing 
({\bf A}) a range of values for  $\alpha_s(Q^2\sim0)$ from 0 to $\infty$~\cite{Deur:2016tte}, and 
({\bf B}) much confusion.
All major nonperturbative approaches to QCD have been solicited\footnote{Except for chiral effective field theory~\cite{Bernard:2006gx}
because it uses hadronic degrees of freedom, which do not couple {\it via} $\alpha_s$.}, namely
LGT, DSE~\cite{Roberts:1994dr, Alkofer:2000wg} and AdS/QCD~\cite{Brodsky:2016yod}, as well as many models.
The different approximations underlying these methods\footnote{{\it e.g.}, spacetime discretization in LGT or truncation prescriptions for the DSE.} 
or the fact that the models are not explicitly rooted in QCD is one reason why their predictions 
differed so much. Other reasons were the differences in the basic definition of $\alpha_s$;
the choice of vertex to compute $\alpha_s$; 
the choice of gauge and RS;
and/or the appearance of multiple solutions to the same equations, providing differing $\alpha_s$~\cite{Fischer:2008uz}.\footnote{
Namely, the {\it scaling}~\cite{Lerche:2002ep} vs. 
{\it decoupling}~\cite{Boucaud:2002fx} solutions.  When $\alpha_s^{gh}$ is computed in the MOM RS
the scaling solution leads to a freezing $\alpha_s^{gh}$ while the decoupling solution leads to a vanishing $\alpha_s^{gh}$. The latter behavior means that if 
a gauge requires ghost fields, 
the ghost stops interacting with the gluons in the IR. 
}
Now, the major nonperturbative methods have converged toward a meaningful definition of $\alpha_s$ that
encompasses IR phenomena~\cite{Deur:2016tte, Deur:2023dzc}. 
Before discussing this advance, we will first briefly present the quest for such a definition, mentioning
only pioneering attempts and glossing over subsequent important works by many, who studied and refined these attempts.

J.~M.~Cornwall, in an influential pioneering work~\cite{Cornwall:1981zr}, developed the {\it pinch technique}  to calculate
using the DSE  
a gauge-independent $\alpha_s$ that freezes in the IR. 
While IR-freezing had been already conjectured at the advent of QCD~\cite{Caswell:1974gg, Sanda:1979xp, Banks:1981nn}, 
it was by no means the only behavior envisioned. Other proposals were that $\alpha_s$ vanishes as 
$Q^2 \to 0$~\cite{Dokshitzer:1995ev, Boucaud:2002fx}, that it diverges as $1/Q^2$~\cite{Richardson:1978bt} 
(based on considering the $\mbox{Q-}\overline{\mbox{Q}}$ static quark potential), 
or that it monotonically rises with $1/Q^2$ without diverging~\cite{Shirkov:1997wi}.
The $\alpha_s$ definition used by Cornwall relies on correlation functions (specifically, the gluon propagator), 
a prevalent way to define $\alpha_s$, see Section~\ref{sec:DSE}. Other notable definitions 
directly use phenomenology,\footnote{For instance, using constituent quark models, the $Q\overline{Q}$ 
potential, or the hadronic spectrum~\cite{Eichten:1974af, Richardson:1978bt, Celmaster:1978jt, Levine:1978rn, Buchmuller:1980bm, Buchmuller:1980su, Godfrey:1985xj}.}
the effective charge concept~\cite{Grunberg:1982fw}, or analytic approaches~\cite{Shirkov:1997wi, Dokshitzer:1995qm}, with the two
latter making $\alpha_s$ an observable.
%

\begin{figure}
\begin{center}
\centerline{\includegraphics[width=0.45\textwidth, height=0.4\textwidth]{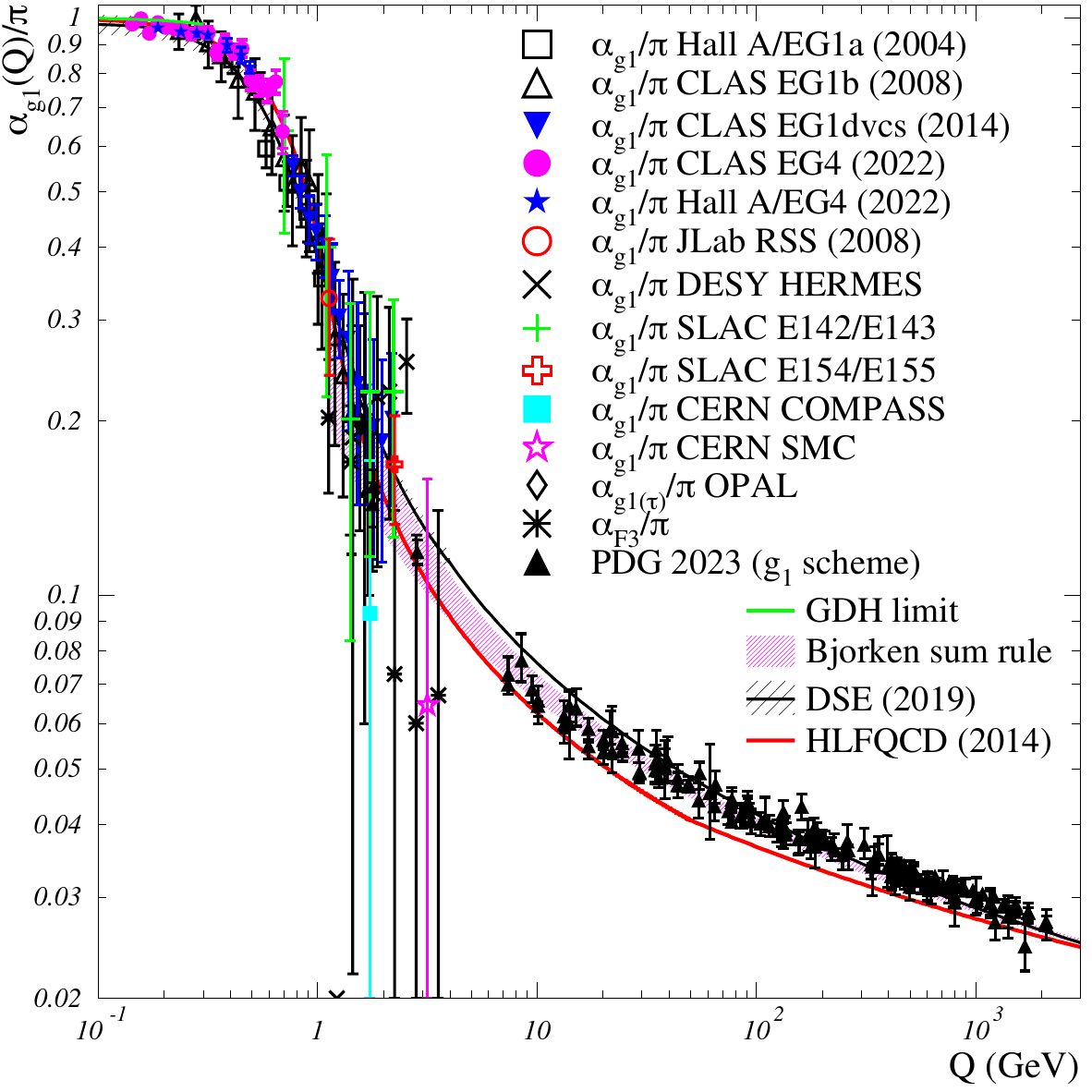}}
\end{center}
\caption{{\bf The QCD coupling at all scales.}  The experimental  data~\cite{Deur:2005cf, Deur:2008rf, Deur:2022msf}, shown by the symbols, are 
in the $g_1$ scheme, {\it viz} the effective charge definition~(Section~\ref{EffectiveCharge}) obtained $via$ the Bjorken sum rule~\cite{Bjorken:1968dy}. 
Exceptions are the OPAL datum  extracted from $\tau$-decay and converted to the $g_1$ scheme using the CSR~\cite{Brodsky:1994eh}, 
$\alpha_{F_3}$ obtained from the GLS sum rule~\cite{Gross:1969jf} and expected to be nearly equal to $\alpha_{g_1}$~\cite{Deur:2016tte}, and the PDG data~\cite{Workman:2022ynf}, converted from $\overline{\rm MS}$ to the $g_1$ scheme.
The nonperturbative calculations use the HLFQCD  framework~(Section~\ref{HLFQCD}), or the DSE formalism with lattice 
determinations of correlation functions (black-hatched band)~(Section~\ref{sec:DSE}).
The green line (top left on the plot) and magenta band are $\alpha_{g_1}$ deduced from, respectively, the 
GDH sum rule~\cite{Gerasimov:1965et, Drell:1966jv} and the Bjorken sum rule together with $\Lambda_s^{\overline{\rm MS}}$ from the PDG.
Other IR determinations of $\alpha_s$ exist, {\it e.g.},~\cite{Cornwall:1981zr} and  \cite{Godfrey:1985xj}. 
When converted into the $g_1$ scheme~\cite{Deur:2016cxb}, they agree with the data.}
\label{fig:alpha_g1}
\end{figure}

From these studies and developments emerged a fruitful definition of the IR coupling, the effective charge method~\cite{Grunberg:1980ja, Grunberg:1982fw, Grunberg:1989xf}.
It was computed~\cite{Binosi:2016nme} using the pinch technique~\cite{Cornwall:1981zr} and the Background Field~\cite{Abbott:1980hw} method. 
The latter enforces gauge independence
and the whole formalism obeys the STI. 
Then, QCD is characterized by a unique $\alpha_s$, the same for all vertices. 
Importantly, and deviating from the original definition~\cite{Grunberg:1982fw}, this $\alpha_s$ is not defined with a specific process 
and therefore, is process-independent. 
It is obtained from correlation functions calculated with either the DSE or LGT~\cite{Binosi:2016nme, Cui:2019dwv}, see Section~\ref{sec:DSE}. 
The result is shown in Fig.~\ref{fig:alpha_g1} and agrees with the phenomenological coupling~\cite{Deur:2005cf, Deur:2008rf, Deur:2022msf} that follows the 
effective charge prescription~\cite{Grunberg:1982fw} applied to the Bjorken sum rule~\cite{Bjorken:1968dy}. It also agrees with
$\alpha_s$ calculated using AdS/QCD~\cite{Brodsky:2010ur}, Section~\ref{HLFQCD}. 
In the next sections, we provide details on the central ingredients mentioned in this brief description, 
 namely effective charges, the AdS/QCD calculation, and the DSE/LGT one.
Other methods that have been used to study $\alpha_s$ in the IR are described in~\cite{Deur:2023dzc, Deur:2016tte, Prosperi:2006hx}

\subsection{Effective charge method}
\label{EffectiveCharge}
QCD effective charges, devised by Grunberg~\cite{Grunberg:1980ja, Grunberg:1982fw, Grunberg:1989xf}, are defined so that pQCD
series stop at first order in their coupling ($\equiv$effective charge) expansion.
Thus, an effective charge includes in the perturbative coupling $\alpha_s^{\rm pQCD}$ all higher-order terms, gluon emission, 
vertex corrections$\cdots$, that drive the higher-order DGLAP evolution. 
We illustrate the concept with the Bjorken sum rule~\cite{Bjorken:1966jh, Bjorken:1969mm,Gribov:1972ri, Kataev:1994gd, Kataev:2005hv, Baikov:2008jh}:  
\begin{align} 
\Gamma_1^{\rm p-n}(Q^2) & \equiv \int_0^1 g_1^{\rm p-n}(x, Q^2) dx
= \frac{g_{\rm A}}{6} \bigg[ 1-\frac{\alpha^{\rm pQCD}_{{\rm s}}(Q^2)}{\pi}
-3.58 \left(\frac{\alpha^{\rm pQCD}_{{\rm s}}(Q^2)}{\pi}\right)^2  
-20.21 \left(\frac{\alpha^{\rm pQCD}_{{\rm s}}(Q^2)}{\pi}\right)^{3}
+ \cdots
\bigg] + \sum_{n > 1} \frac{\mu_{2n}(Q^2)}{Q^{2n-2}},
 \label{bjorken SR}
\end{align} 
with $g_1^{\rm p-n}(x, Q^2)$ the isovector 
nucleon spin structure function, 
$x$ the Bjorken scaling variable, 
$g_A$ the nucleon axial charge and $\mu_{2n}(Q^2)$ are higher-twist coefficients.
The series coefficients are in $\overline {\rm MS}$ and for $n_f = 3$. 
The associated effective charge $\alpha_{g_1}$ is then defined by:
\begin{align}
\Gamma_1^{\rm p-n}(Q^2)  \equiv \frac{g_A}{6}\bigg(1-\frac{\alpha_{g_1}(Q^2)}{\pi}\bigg) 
~~~\Longrightarrow~~~ \alpha_{g_1}(Q^2)  = \pi \left(1-\frac{6}{g_A}\Gamma_1^{\rm p-n}(Q^2) \right)
\label{eqn:alphadef}
\end{align}
where the label $g_1$  in $ \alpha_{g_1}$ indicates the observable defining the effective charge. 
Hence, both short distance (bracket in Eq.~\eqref{bjorken SR}) and long distance ($\mu_{2n}/Q^{2n-2}$ terms) effects became encapsulated in $\alpha_{g_1}$. 
This generalizes the RGE procedure that makes the coupling {\it constant} to run by including 
quantum loop effects in it. Effective charges remain defined in the IR and
are RS-independent since any pQCD approximant is RS-independent at LO (Section~\ref{alpha-s in the UV}). In fact,
they are observables, thus devoid of Landau p\^ole. 
In the UV, Eqs.~(\ref{bjorken SR}-\ref{eqn:alphadef}) yield the relation between $\alpha_{g_1}$ and the standard pQCD coupling:
{\allowdisplaybreaks
\begin{eqnarray}
\alpha_{g_1}(Q^2) = \alpha^{\rm pQCD}_s(Q^2) + 3.58\frac{({\scriptstyle \alpha^{\rm pQCD}_s(Q^2)})^2}{\pi} +
20.21\frac{({\scriptstyle \alpha^{\rm pQCD}_s(Q^2)})^{3}}{\pi^2} + \cdots ~~~\biggr\rvert_{\rm UV~only}
\label{eq:msbar to g_1}
\end{eqnarray}
}
More generally, once an observable 
is chosen to define an effective charge,
all other effective charges follow {\it via} commensurate scale relations (CSR)~\cite{Brodsky:1994eh}, thereby ensuring the predictive power 
of QCD.\footnote{
CSR are known in the UV domain. Extending them to the IR is discussed in Refs.~\cite{Brodsky:1994eh, Deur:2014qfa, Deur:2016cxb, Deur:2016opc}.}
CSR also show that the choice of observable defining an effective charges is equivalent to a RS choice.

It is advantageous to define an effective charge with the Bjorken sum rule because
its pQCD series is relatively simple and known to high order (N$^4$LO). Furthermore, $g_A$ is well-known~\cite{Workman:2022ynf},
$\Gamma_1^{\rm p-n}(Q^2)$ is extensively measured and where measurements are lacking, reliable 
relations~\cite{Gerasimov:1965et, Drell:1966jv, Bjorken:1966jh, Bjorken:1969mm} supplement them~\cite{Deur:2023dzc}. But most importantly, $\alpha_{g_1}$ is, 
to good approximation, interpretable as a standard coupling, including in the IR.\footnote{
Most other effective charges have complicated interpretations, if any, and behaviors much different than expected from a force coupling: some, {\it e.g.}, are
negative.} 
This is because $\Gamma_1^{\rm p-n}$ is an isovector integral in which 
coherent effects (i.e., involving rigidly linked quarks) are highly 
suppressed~\cite{Deur:2023dzc, Deur:2016tte, Deur:2009zy}. 
This is remarkable since in the IR, 
individual quarks cannot usually be isolated by the probing process, 
making a single quark's interaction inaccessible, while it is what is needed to  measure the force magnitude. Fig.~\ref{fig:alpha_g1} shows
$\alpha_{g_1}$ extracted from the world data on $\Gamma_1^{\rm p-n}$~\cite{Deur:2014vea, Deur:2021klh, Ackerstaff:1997ws, Ackerstaff:1998ja, 
Airapetian:1998wi, Airapetian:2002rw, Airapetian:2006vy, Kim:1998kia, Adeva:1993km, Alexakhin:2006oza, Alekseev:2010hc, Adolph:2015saz, 
Anthony:1993uf, Abe:1994cp, Abe:1995mt, Abe:1995dc, Abe:1995rn, Anthony:1996mw, Abe:1997cx, Abe:1997qk, Abe:1997dp, Abe:1998wq, 
Anthony:1999py, Anthony:1999rm, Anthony:2000fn, Anthony:2002hy}.

\subsection{Holographic Light-Front  QCD }
\label{HLFQCD}
The AdS/QCD calculation of $\alpha_s$ is done with the HLFQCD model~\cite{Brodsky:2014yha}, an approach based on
light-front (LF) canonical quantization
, in which a field is quantized using LF time, $x^+ \equiv  t+z$,
(here, $z$ is one of the 3 space coordinates) rather than the usual Galilean time $t$~\cite{Dirac:1949cp, Brodsky:1997de}
. This results in a
Poincar\'e-invariant formalism, free of the pseudo-dynamics that complicates the usual canonical quantization based on
$t$~\cite{Brodsky:2022fqy, Deur:2024unruh}. LF quantization provides a rigorous nonperturbative approach to solving  QCD, resulting, for $m_q \to 0$, in
hadronic structures described by a relativistic Schr\"{o}dinger equation~\cite{deTeramond:2008ht}. 
The equation is solvable but the confining potential term has thus far been too difficult to compute ab-initio. Instead,  
gauge-gravity, or AdS/CFT, duality~\cite{Maldacena:1997re} 
is used. It posits that classical gravity in a $(n+1)$D negatively curved spacetime (anti-de Sitter, AdS, spacetime) is dual
to a conformal field theory (CFT) residing on the boundary of the $(n+1)$D spacetime, {\it i.e.}, in a $n$D Minkowski (flat) spacetime.
Crucially, for $m_q \to 0 $, QCD's Lagrangian, Eq.~\eqref{eq:QCD Lagrangian}, is that of a CFT.
We saw in Section~\ref{alpha-s in the UV}, that the classical conformal symmetry is broken by a quantum anomaly that manifests as
a $\sim$1~GeV phenomenological scale. Far enough from that scale, QCD is approximately conformal: deep-UV displays
Bjorken scaling~\cite{Bjorken:1969ja, Feynman:1969wa}, and a similar dearth of $Q^2$-dependence is observed in the deep-IR: $\alpha_s$ freezes. 
Then, AdS/CFT applied to QCD yields AdS/QCD.\footnote{
Formally, the AdS$\leftrightarrow$CFT duality stems from the similitude between the group of isometries in 5D AdS and the SO(4,2) conformal group
describing QCD if $m_q \to 0 $.}
A semiclassical\footnote{Since the gravity theory is classical, HLFQCD is a semiclassical approximation to QCD.
This makes HLFQCD strictly valid only in the IR, since in the UV,  $\alpha_s$ runs due to
quantum loops. Yet, we will see that HLFQCD can be smoothly merged with pQCD, providing a coupling valid at all $Q^2$.}
 potential for the relativistic Schr\"{o}dinger equation can then be computed using generic symmetries of QCD for $m_q \to 0 $. 
Several ways exist to do so, all leading to the same harmonic oscillator form, $U(\zeta)= \kappa^4 \zeta^2 + b$,
where $\zeta$ is the transverse parton separation in LF coordinates, $\kappa$ is the quantum anomaly's scale (therefore directly related to
$\Lambda_s$~\cite{Deur:2014qfa, Deur:2016cxb, Deur:2016opc}), and $b$ depends on $\kappa$ and the spin representations in AdS space.
For $m_q=0$, $\kappa$ is universal. It can be deduced from any hadron masses, {\it e.g.}, $\kappa=M_{\rm p}/2$, 
with $M_{\rm p}$ 
the proton mass~\cite{Brodsky:2014yha}, or from 
$\Lambda_s$~\cite{Deur:2016opc, deTeramond:2024ikl}.
To the $U(\zeta)$ in the LF theory corresponds a deformation of the AdS space given by a term $\exp(\kappa^2 z_h^2)$ factoring the AdS metric, 
with $z_h$ being the $5^{\rm th}$ (``holographic'') coordinate of the AdS space. The factor $\exp(\kappa^2 z_h^2)$ breaks conformality and grows with 
$z_h$, causing confinement. Then, $z_h$ is related to the CFT momentum scale~\cite{Peet:1998wn}
: $Q \sim 1/z_h$.

To calculate $\alpha_s$ in HLFQCD, one starts with the general relativity Action applied to AdS spacetime: 
\begin{equation} 
S = -\frac{1}{4} \int \sqrt{g}\frac{1}{\hat{a}^2}F^2~d^5x,
\end{equation} 
where $F$ is the field, $\hat{a}$ its coupling and $g \equiv \det(g_{\mu \nu})$. $g_{\mu \nu}$ is the AdS metric of invariant interval 
$ds^2$=$\frac{R^2}{z_h^2} \big(\eta_{\mu \nu} dx^\mu dx^\nu$-$dz_h^2 \big)$, with
$R$ the AdS radius and $\eta_{\mu \nu}$, the Minkowski metric. As said, this is dual to a CFT. To break conformal invariance 
while providing the mandatory harmonic oscillator form for $U(\zeta)$, a term $e^{\kappa^2 z_h^2}$ warping the AdS geometry is factored:  
$ds^2$=$\frac{R^2}{z_h^2} e^{\kappa^2 z_h^2}\big(\eta_{\mu \nu} dx^\mu dx^\nu$-$dz_h^2 \big)$. The action then becomes:
\begin{equation} 
S_{HLF}=  -\frac{1}{4} \int \sqrt{g}\frac{1}{{\hat{a}}^2}F^2 ~e^{\kappa^2 z_h^2}d^5x.
\label{eq:AdS distorted action}
\end{equation} 
As explained in Section~\ref{EffectiveCharge}, effective charges generalize the RGE running couplings by encapsulating, in addition to quantum loops, gluon radiation (higher-order DGLAP corrections), parton distribution correlations (higher-twists), and long-distance confinement effects.
Accordingly, in Eq.~\eqref{eq:AdS distorted action}, the $e^{\kappa^2 z_h^2}$ confinement term is attached to
the coupling: $a(z_h^2) \equiv {\hat{a}} e^{-\kappa^2 z_h^2/2}$~\cite{Pirner:2009gr, Brodsky:2010ur, Gursoy:2007cb, Gursoy:2007er}. 
Transforming $a(z_h^2)$ to Minkowski 4-momentum space gives $\alpha^{\rm HLF}_s(Q^2) = \alpha^{\rm HLF}_s(0) e^{Q^2/[4\kappa^2]}$.
The normalization $\alpha^{\rm HLF}_s(0)$ is not provided by HLFQCD but
determined by the scheme choice, {\it e.g.}, the $g_1$ scheme~\cite{Brodsky:1994eh} imposes:
\begin{equation} 
\alpha^{\rm HLF}_{g_1}(Q^2) = \pi {\rm e}^{-Q^2/[4\kappa^2]}.
\label{alpha_s from HLFQCD}
\end{equation} 
The normalization $\alpha^{\rm HLF}_s(0)$ marks the effective charge dependence on an observable, similar to the RS-dependence of $\alpha^{\rm pQCD}_s$.
The prediction, Eq.~(\ref{alpha_s from HLFQCD}), has no free parameters and agrees well with data in the IR where HLFQCD is valid (Fig.~\ref{fig:alpha_g1}). 

One can interpolate between the IR Gaussian form, Eq.~(\ref{alpha_s from HLFQCD}) and the UV $\log$ behavior, Eq.~(\ref{Eq.one loop}),
with the form~\cite{deTeramond:2024ikl}:
\begin{equation} 
\alpha^{\rm HLF}_{g_1} = \pi \exp\left[-\int_0^{Q^2} \frac{d u}{4 \kappa^2 + u \ln\left( u/\Lambda_s^2 \right)} \right].
\label{alpha_s HLFQCD-2}
\end{equation} 
Analytic continuation in the complex $Q^2$-plane removes pQCD's Landau p\^ole and, crucially, links 
$\kappa$ to $\Lambda_s$.  A simple $Q^2$-dependence for $\kappa$ determined by the meson spectrum, accounts 
for $m_q \neq 0$~\cite{Dosch:2025}. The result is an analytic description of $\alpha_s$ at any
$Q^2$ that agrees with IR and UV data, {\it viz} over more than 8 orders of magnitude: $2\times10^{-2}<Q^2<4.4\times10^{-6}$~GeV~\cite{Dosch:2025}.
What happens to the Landau p\^ole is revealing: the p\^ole in Eq.~(\ref{alpha_s HLFQCD-2}) obeys
$4 \kappa^2 + Q^2 \ln \left(\sfrac{Q^2}{\Lambda_s^2}\right) = 0$. Continuation in the $Q^2$-plane\footnote{
The continuation is permitted because $\alpha_{g_1}$ is an effective charge, thus an observable.} 
shows that the Landau p\^ole (pQCD case, corresponding to $\kappa=0$) on the real axis migrates to the imaginary $Q^2$ axis 
as $\kappa$ increases. It reaches the imaginary axis ({\it maximum analyticity}) for
\begin{equation}
\kappa=\sqrt{2\pi}\Lambda_s/4. 
\label{eq:maxana}
\end{equation}  
The real-axis unphysical p\^ole has become an imaginary-axis physical feature exposing color confinement. 
From $\kappa= M_{\rm p}/2$ and Eq.~(\ref{eq:maxana}), 
$\Lambda_s^{g_1}=\sqrt{2/\pi}M_{\rm p}$ in the $g_1$ scheme, which becomes\footnote{Eq.~(3.39) of Ref.~\cite{Deur:2023dzc},
with $v_2=-3.58/\pi$ for $n_f=3$, see Eq~(\ref{eq:msbar to g_1}).}  
in $\overline{\rm MS}$ $\Lambda_s^{\overline{\rm MS}}=\sqrt{2/\pi}M_{\rm p}e^{-\frac{2}{\beta_0}3.58} =
0.34$~GeV for $n_f=3$, agreeing  with the world data~\cite{Workman:2022ynf}.

The Landau p\^ole metamorphosis from an unphysical feature to a manifestation of color confinement is naturally interpreted: imaginary p\^oles 
reveal dissipative/irreversible effects that suppress the time evolution of processes. For instance, the diverging oscillations of a driven harmonic oscillator 
are dissipated by friction formalized by an imaginary p\^ole in frequency space~\cite{Arfken-Weber}. Oscillation dissipation suppresses 
the time evolution, {\it viz}, the propagation, of the system. The larger the p\^ole's imaginary value, the larger the dissipation and, thus, 
the more the propagation is suppressed. Since $\alpha_s$ is a product of parton propagators; {\it e.g.}, 
Eq.~(\ref{eq:alpha_s DSE ghost--gluon}), its p\^ole reflects p\^oles in parton propagators. Then, as the Landau p\^ole moves from real 
to complex to imaginary values, it suppresses parton propagation, eventually confining them (full propagation suppression) 
at the maximum analyticity condition.

\subsection{Dyson-Schwinger equations \label{sec:DSE}}

The DSE~\cite{Dyson:1949ha, Schwinger:1951ex, Schwinger:1951hq} are another approach providing $\alpha_s$. The DSE 
are the equations of motion of QCD (or any QFT) and rigorously provide its correlation functions, {\it viz} propagators and 
vertex functions. The DSE generate an infinite set of coupled non-linear integral equations, with the equation for the $n$-point 
function coupled to the equations for the $n$+1 or $n$+2-point functions. 
In principle the DSE provide exact solutions but in practice, the infinite equation set must be limited (``truncated''). This must be done carefully, 
lest it creates unphysical artifacts. While this was historically a delicate issue, recent progress has identified symmetry-preserving  schemes with 
controlled truncation-dependent uncertainties.
The first DSE calculation of a QCD running coupling, viz. in a framework including quarks is discussed in~\cite{Fischer:2003rp}.
To compute $\alpha_s$ with the DSE, one expresses it {\it via} renormalization ``constants'', see
Refs.~\cite{Muta:1998vi, Alkofer:2004it, Deur:2023dzc} for 1-loop UV computations of $\alpha_s$. 
Consider $Z_\alpha(Q^2,\mu^2)$, the function that evolves $\alpha_s$ from an arbitrarily chosen renormalization scale $\mu$
to another scale $Q^2$, {\it viz} $Z_\alpha(Q^2,\mu^2) \equiv \alpha_s(\mu^2)/ \alpha_s(Q^2)$. 
For,  {\it e.g.}, the ghost-gluon vertex, $Z_\alpha=\tilde{Z}_1^2/(\tilde{Z}_3^2 Z_3)$, 
where, $Z_3$, ${\tilde{Z}_1}$, and $\tilde{Z}_3$ are the renormalization constants of the gluon propagator,
ghost-gluon vertex, 
and ghost propagator, 
respectively. 
Computing $\alpha_s$ then amounts to computing the relevant $Zs$. Here, $Z_\alpha=\tilde{Z}_1^2/(\tilde{Z}_3^2 Z_3)$ 
provides $\alpha_s^{gh}$, the ghost-gluon coupling\footnote{Also called {\it Taylor coupling}~\cite{Taylor:1971ff}. Being the easiest coupling that can be computed using correlation functions, it is prominent in the literature.} (Fig.~\ref{Fig:various_vertex_couplings}):
\begin{equation}
\alpha_s^{\rm gh}\left(Q^2\right)= \alpha_s^{\rm gh} \left(\mu\right)G^2
\left(Q^2,\mu\right)Z\left(Q^2,\mu\right),
\label{eq:alpha_s DSE ghost--gluon}
\end{equation}
where $G(Q^2,\mu)$ and $Z(Q^2,\mu)$ are respectively the ghost and gluon propagator dressing functions, which are calculable using the DSE.\footnote{
They are also conveniently calculated with LGT, the Functional Renormalization Group, pQCD (UV only), and other methods, see~\cite{Deur:2023dzc}.}
Other vertices and associated renormalization constants can be used, yielding the  
3-gluon coupling ($\alpha_s^{\rm 3g}$), 4-gluon coupling ($\alpha_s^{\rm 4g}$), and gluon-quark coupling ($\alpha_s^{\rm gq}$), 
see~\cite{Deur:2023dzc} for their formulae. Couplings can also be defined using single propagators, {\it e.g.}, the quark one~\cite{Cornwall:1981zr}.
These definitions yield couplings that typically differ in the IR. Even after selecting a vertex, couplings may still differ
due to gauge and kinematic choices ({\it viz}, what parton momentum-flow is considered~\cite{Deur:2023dzc}).
We remind that while in the UV, and for RS independent of $m_q$, all couplings are identical, it is not necessarily true in the IR. 
We show here an example of formalism that does provides a single IR coupling, interpretable as a
vertex/process-independent effective charge, thus comparable to $\alpha_{g_1}$, Eq.~(\ref{eqn:alphadef}). 
This was achieved~\cite{Binosi:2016nme, Cui:2019dwv} by making QCD's correlation functions to retrieve some of QED's abelian features.\footnote{
By systematically rearranging classes of diagrams so that their sums result in correlation functions
that obey the STI.} Then, like for QED, vacuum polarization loops drive the running, consistent with the RGE.
The resulting effective gluon dressing function $ {\mathcal D}(Q^2)$ provides a universal, process-independent and effectively gauge-independent coupling $\hat\alpha_s$~\cite{Binosi:2016nme, Cui:2019dwv}: 
\begin{equation}
\label{alphahatequation}
\hat\alpha_s(Q^2) {\mathcal D}(Q^2) = \alpha_0 
\left[\frac{G(Q^2;\mu^2)/G(0;\mu^2)}{1-L(Q^2;\mu^2)G(Q^2;\mu^2)}\right]^2
{\mathsf D}(Q^2)
~,
\end{equation}
with $L(Q^2;\mu^2)$ the longitudinal component of the gluon-ghost vacuum polarization that vanishes at $Q^2=0$ and 
${\mathsf D}(Q^2) \equiv \frac{Z(Q^2,\mu)}{Q^2} \frac{m_g^2(Q^2,\mu)}{m_0^2}$. 
The factor $m_g(Q^2,\mu)$ is interpreted as an effective running gluon mass that 
originates~\cite{Binosi:2022djx, Papavassiliou:2022wrb, Aguilar:2022thg} from the Schwinger mechanism~\cite{Schwinger:1962tn, Schwinger:1962tp}.
Its magnitude is set by the IR limit $m_g \to m_0$, while in the UV, $m_g \to 0$ as it must.
Then, ${\mathcal D}(Q^2)$ behaves as the propagator of a free, effectively massive, dressed gluon.
The $m_g(Q^2,\mu)$ causes $\hat\alpha_s$ to lose its $Q^2$-dependence in the IR, freezing at $\hat \alpha_0=0.97(4)\pi$, 
since for $Q \lesssim m_0$, the relevant scale becomes $m_0$ rather than $Q$.  
%
Following the concept of effective charge, Eq.~\eqref{alphahatequation} inserts in the coupling definition a renormalization group invariant interaction. In fact, the LHS of Eq.~\eqref{alphahatequation} 
expresses a force: {$F$=(coupling$\times$propagator)}. Hence, it incorporates color confinement~\cite{Binosi:2014aea}  like
$\alpha_{g_1}$ and can be compared to it. The ingredients of Eq.~\eqref{alphahatequation} were computed~\cite{Cui:2019dwv} by combining DSE results verified by LGT and, notwithstanding that $\hat\alpha_s(Q^2)$ has no adjustable parameters, it agrees well with $\alpha_{g_1}$ (Fig.~\ref{fig:alpha_g1}).
%

\subsection{Summary and Perspective}
\label{epilogue}

The QCD coupling $\alpha_s$ is a central ingredient of QCD and more generally, of the Standard Model. 
To understand it, vigorous efforts are ongoing on two separate fronts.

The UV front is the pursuit of an accurate determination of $\alpha_s(M_Z)$. The techniques involved,
the RGE 
and perturbation theory, are well-known. 
They yield an $\alpha_s(Q^2)$ that, at first order, logarithmically decreases with $Q^2$. 
Thus, both $\alpha_s$ and its $Q^2$-dependence vanish as $Q^2 \to \infty$, allowing to use pQCD to predict high-energy reactions, and 
proving that QCD is the correct gauge theory of the strong force. The current goal on the UV front is to determine $\alpha_s(M_Z)$ to 
well below the sub-percent accuracy~\cite{dEnterria:2022hzv}. 
Presently (2024), $\Delta \alpha_s/\alpha_s=0.76\%$~\cite{Workman:2022ynf}. This still makes $\Delta \alpha_s$ to
contribute notably to uncertainties in pQCD calculations. Hence, hadronic uncertainties often dominate in calculations of Standard Model reactions,
hindering not only studies of QCD and the Standard Model, but also searches for new physics.
Reducing $\Delta \alpha_s$ necessitates combining many determinations of $\alpha_s(M_Z)$ both from distinct experiments and from
LGT. It also requires improving our knowledge of pQCD approximants to higher orders, and pursuing techniques that minimize pQCD ambiguities~\cite{Brodsky:1982gc,Brodsky:2011ta}. 
To reach a satisfactory $\Delta \alpha_s$ will be a long quest, since even the sub-percent goal is far from our knowledge of the
other fundamental couplings.\footnote{
$\Delta \alpha/\alpha=1.5\times 10^{-10}$,  
$\Delta \alpha_w/\alpha_w=5.1\times 10^{-7}$  
and $\Delta G_N/G_N=2.2\times 10^{-5}$ 
\cite{Workman:2022ynf}.}

The IR front is the study of $\alpha_s$ in the strongly-coupled QCD regime. Its challenges differ from those of the UV.
{\bf Firstly}, different definitions are available for $\alpha_s$.
{\bf Secondly}, multiple couplings are often necessary to characterize QCD following how its different vertices couple. 
Such couplings tell us how quark, gluon and, if required, ghost fields interact at low energy in a given RS and gauge, 
Yet, they do not directly reflect QCD's strength, which is what is usually expected from a fundamental coupling. This particular hurdle is cleared by either 
making $\alpha_s$ an observable~\cite{Grunberg:1982fw} or enforcing the Slavnov-Taylor 
identities~\cite{Taylor:1971ff, Slavnov:1972fg}, QCD's equivalent of QED's Ward identities~\cite{Ward:1950xp, Takahashi:1957xn}. 
While the former also ensures that $\alpha_s$ is finite and independent of RS and gauge, the latter usually does not. 
{\bf Thirdly}, controlling the uncertainties of nonperturbative calculations or models is difficult. 
It has led to erroneous $\alpha_s$ behaviors arising from artifacts attached to under-controlled approximations. 
%
These challenges resulted in $\alpha_s$ predictions ranging from vanishing to diverging~\cite{Deur:2016tte},
hampering consensus on the IR behavior of $\alpha_s$. 
To progress, a definition of $\alpha_s$ must be identified
that provides a single, gauge-independent, RGE-compliant coupling whose meaning is clear, 
 whose value reveals the quark-quark interaction strength at that $Q^2$, and that is directly comparable to data.   
QED does this by defining $\alpha$ as an effective charge~\cite{GellMann:1954fq}, which
suggests to do the same for QCD~\cite{Grunberg:1982fw, Brodsky:2010ur, Binosi:2016nme}. 
Following this path produced
not only consistent calculations for $\alpha_s(Q^2)$ once the RS-dependence is accounted for~\cite{Deur:2016tte}, but also 
agreement with the experimental data 
(Fig.~\ref{fig:alpha_g1}). 
Calculations and data reveal that $\alpha_s$ plateaus for $Q^2 \ll \Lambda^2_s$.
Another character that the definition must display is practical usefulness: hadronic quantities must be calculable with the coupling. 
The aforementioned definition fulfills this as it enters the calculations of hadron masses, unpolarized, polarized and generalized parton distributions, 
form factors, meson decay constants, and characteristic scales such as $\Lambda_s$~\cite{Deur:2014qfa, Chang:2011ei, 
deTeramond:2018ecg, deTeramond:2021lxc, Chang:2013pq, Shi:2015esa, Ding:2015rkn, Ding:2019qlr, Ding:2019lwe, Yin:2023dbw, 
Sufian:2016hwn, Raya:2015gva, Raya:2016yuj, Rodriguez-Quintero:2018wma, Xu:2022kng, Deur:2016opc, Dosch:2022mop}.
Overall, 
we now have a compelling candidate for an $\alpha_s$ that universally characterizes QCD's strength, that is gauge-, vertex-, and process-independent, and that can predict a wide range of nonperturbative hadronic quantities.


\begin{ack}[Acknowledgments]%
This work is supported by the U.S.\ Department of Energy, Office of Science, Office of Nuclear Physics, contract DE-AC05-06OR23177
\end{ack}

\bibliographystyle{Numbered-Style} 
\bibliography{BibAlpha_EPP}

\end{document}